\documentclass[pra,aps,twocolumn,10pt,floatfix,longbibliography,superscriptaddress]{revtex4-1}

\usepackage{pdfpages}
\usepackage{etoolbox} 

\usepackage{soul}
\usepackage{amsmath}
\usepackage{amsfonts}
\usepackage{txfonts}
\usepackage{amssymb}
\usepackage{xr-hyper}
\usepackage{hyperref}
\hypersetup{colorlinks=true, linkcolor=blue, citecolor=blue, urlcolor=blue}
\usepackage[capitalize]{cleveref}
\usepackage{graphicx}
\usepackage[justification=Justified,format=plain,singlelinecheck=false]{caption}
\usepackage{subcaption}
\usepackage{bbold}					
\usepackage[makeroom]{cancel}		
\usepackage{multirow}				
\usepackage[normalem]{ulem}         
\usepackage{array}
\usepackage{mathrsfs}
\usepackage{mathtools}
\usepackage[version=4]{mhchem}      
\usepackage{IEEEtrantools}          
\usepackage{dcolumn}                
\usepackage{physics}                
\usepackage{dsfont}                 
\usepackage{enumerate}              
\usepackage[shortlabels]{enumitem}  
\usepackage{verbatim}
\usepackage{csquotes}
\usepackage{pifont}

\usepackage{mathdots}

\newcommand{\beq}[1]{\begin{equation}\label{#1}}
\newcommand{\eep}{\;.\end{equation}}
\newcommand{\eec}{\;,\end{equation}}
\newcommand{\eeq}{\end{equation}}

\renewcommand{\d}{\delta}
\newcommand{\s}{\sigma}

\newcommand{\om}{\omega}

\newcommand*\chem[1]{\ensuremath{\mathrm{#1}}} 

\newcommand{\Ct}{\mathcal{C}_3}

\DeclareMathAlphabet{\mathcal}{OMS}{cmsy}{m}{n} 

\let\SectionOriginal\section
\newcommand*{\headingSectionPRL}[1]{\belowpdfbookmark{#1}{#1}{\textit{#1.---}}\ignorespaces}
\let\section\headingSectionPRL

\renewcommand{\vec}[1]{{\bf #1}}

\newcommand{\kv}{\vec{k}}
\newcommand{\xv}{\vec{x}}
\newcommand{\rv}{\vec{r}}

\newcommand{\Pv}{\vec{P}}

\newcommand{\intBZ}{\int_{\text{BZ}}} 

\newcommand{\sv}{\boldsymbol\sigma} 

\begin{document}

\makeatletter
\patchcmd{\@outputpage@head}{\@ifx{\LS@rot\@undefined}{}{\LS@rot}}{}{}{}
\makeatother

\title{Shift photocurrent vortices from topological polarization textures}

\date{\today}

\newcommand{\TCM}{{Theory of Condensed Matter Group, Cavendish Laboratory, University of Cambridge, J.\,J.\,Thomson Avenue, Cambridge CB3 0HE, UK}}
\newcommand{\HarvardSeas}{John A.~Paulson School of Engineering and Applied Sciences, Harvard University, Cambridge, Massachusetts 02138, USA}


\author{Aneesh Agarwal}
\email{aa2223@cam.ac.uk}
\affiliation{\TCM}

\author{Wojciech J. Jankowski}
\affiliation{\TCM}

\author{Daniel Bennett}
\affiliation{\HarvardSeas}

\author{Robert-Jan Slager}
\email{rjs269@cam.ac.uk}
\affiliation{\TCM}
\affiliation{Department of Physics and Astronomy, University of Manchester, Oxford Road, Manchester M13 9PL, UK}

\begin{abstract}
Following the recent interest in van der Waals (vdW) ferroelectrics, topologically nontrivial polar structures have been predicted to form in twisted bilayers. However, these structures have proven difficult to observe experimentally. We propose that these textures may be probed optically by showing that topological polarization textures result in exotic nonlinear optical responses.
We derive this relationship analytically using non-Abelian Berry connections and a quantum-geometric framework, supported by tight-binding and first-principles calculations.
For the case of moir\'e materials without centrosymmetry, which form networks of polar merons and antimerons, the shift photoconductivity forms a vortex-like structure in real space.
For a range of frequencies where transitions between topologically trivial bands occur at the Brillouin zone edge, the shift photocurrents are antiparallel to the in-plane electronic polarization field.
Our findings highlight the interplay between complex polarization textures and nonlinear optical responses in vdW materials and provide a sought-after strategy for their experimental detection.

\end{abstract}

\maketitle

\section{Introduction}
Twisting layered van der Waals (vdW) materials to form interference patterns known as moir\'e superlattices \cite{Bistritzer2011,carr2017twistronics} offers a broad platform for realizing exotic physical phenomena, including superconductivity \cite{cao2018unconventional}, correlated \cite{nuckolls2020strongly, wu2021chern} and fractional \cite{xie2021fractional,zeng2023thermodynamic,park2023observation} Chern insulators, and the appearance and manipulation of magnetic \cite{tong2018skyrmions,hejazi2020noncollinear,song2021direct,bennett2024stacking} and polar \cite{stern2020interfacial,yasuda2021stacking,wang2022interfacial,weston2022interfacial,ko2023operando,molino2023ferroelectric,van2024engineering} order in two-dimensional (2D) systems.
In particular, stacking-engineering of vdW materials has been shown to result in ferroelectricity with state-of-the-art performance in nanoscale devices \cite{yasuda2021stacking,yasuda2024ultrafast,bian2024developing}.
Introducing a relative twist in vdW ferroelectrics results in the formation of a regular network of moir\'e polar domains (MPDs) \cite{ko2023operando,molino2023ferroelectric,van2024engineering}, the origin of which has been attributed to the symmetry-breaking of the different stacking arrangmenents in a moir\'e superlattice \cite{li2017binary,bennett2022electrically,bennett2022theory}.

This symmetry breaking also gives rise to in-plane polarization textures in the MPDs, giving them topological character \cite{bennett2023polar,bennett2023theory} and providing a new platform to induce band topology~\cite{Rmp1,Rmp2}. 
In this regard it was recently shown that the real space topology of polar textures is compatible with non-trivial band topology \cite{jankowski2024polarization}.
While similar polar topological textures have been observed in oxide perovskites \cite{das2019observation, han2022high, junquera2023topologicaly, sanchez20242d}, this is the first such prediction in vdW materials, and in a 2D system (less than 1~nm thick).
The topological character of the MPDs has recently been confirmed in twisted \chem{WSe_2} using piezoresponse force microscopy (PFM) \cite{vu2024imaging}.
Understanding the physical consequences of polar topological structures in moir\'e materials, namely how they influence other materials properties, may lead to advances in nanotechnology, and may also reveal new ways to detect these exotic structures experimentally. 
The advancement of this new direction in nanotechnology hinges on identifying physical observables to harbor and manipulate these exotic states.

\begin{figure}[t!]
\includegraphics[width=\columnwidth]{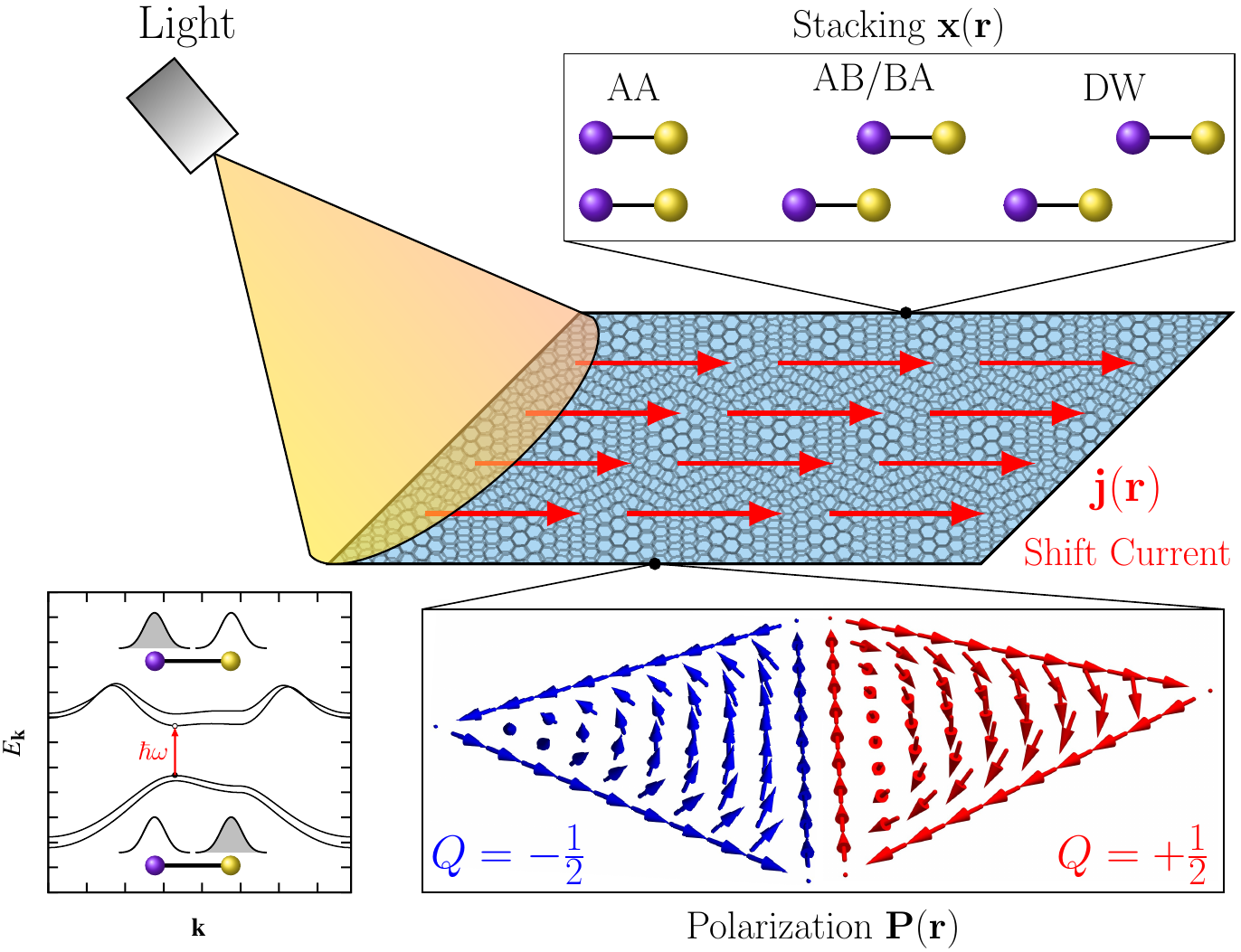}
\caption{
{\bf Photocurrent response from polarization textures.}
A polar moir\'e material is shown, in which a network of stacking domains forms.
The high symmetry stackings, AA, AB/BA and DW are sketched above.
The stacking domains have topologically nontrivial polarization, forming a network of merons and antimerons (winding numbers $Q=\pm\frac{1}{2}$), sketched below.
Illuminating the sample results in an inhomogeneous shift photocurrent, caused by the shift of Wannier centers between the valence and conduction bands, sketched below.
}
\label{Fig1}
\end{figure}

Optical measurements constitute one of the key tools for experimentally probing new physical effects, with many intriguing optical signatures~\cite{Belinicher1982} being reported, most recently in moir\'e materials~\cite{ochoa2020, Hesp_2021, Zhang_2023, Du2023, kuang2024opticalpropertiesplasmonsmoire,zhang2024plasmonic}. 
In particular, nonlinear optical effects~\cite{Sipe1993, Sipe2000} such as the shift response~\cite{Belinicher1982, Sipe1993, chaudhary2022, chen2024enhancingshiftcurrentvirtual}, which yield polarization currents due to photoinduced shifts of electronic charge centers, can result in bulk DC responses: 
the generation of these DC currents from light-induced excitations is of interest for photovoltaic applications~\cite{Cook2017,kaplan2022twisted,zhu2024anomalousshiftopticalvorticity}. 
The theory of optical responses of topological states has recently renewed~\cite{bouhon2023quantum, tormaessay,Ahn2021} interest in relations with quantum geometry~\cite{provost1980riemannian}. 
The geometry of quantum states can be described in terms of the quantum geometric tensor (QGT) that encodes non-Abelian, i.e., multiband, Berry connections~\cite{vanderbilt2018berry} and their derivatives~\cite{provost1980riemannian}.
Apart from many relations to a diverse set of physical observables that range from superfluid densities to wavefucntion spreading, the QGT describes dipole transitions and hence is a useful quantity for capturing topological optical responses at linear and non-linear orders~\cite{Ahn2021, bouhon2023quantum}.

In this work, we uncover an intriguing interplay between topological polar structures~\cite{bennett2023polar, bennett2023theory} and the optical and geometric properties of moir\'e materials~\cite{topp2021} that host topologically trivial bands, see Fig.~\ref{Fig1}.
We discover a definite optical signature for the topological character of polar domains in moir\'e materials in the absence of centrosymmetry. Our findings can be described with a quantum-geometrical framework~\cite{Ahn2021}, using its relation to the polarization carried by the Wannier charge centers~\cite{Sipe1993}, see Fig.~\ref{Fig1}.
We illustrate these findings using tight-binding and first-principles calculations, using bilayer hexagonal boron nitride (hBN) as an example of a prototypical vdW ferroelectric, although our results are applicable to a wide variety of vdW materials such as transition metal dichalcogenides (TMDs).

\section{Results}
The electronic polarization in a 2D crystal is given by \cite{king1993theory, vanderbilt1993electric, resta1994macroscopic}
\beq{eq:KSV}
 \vec{P} = -\frac{e}{(2 \pi)^2} \intBZ \dd^2 \kv~ \sum^{\text{occ}}_n f_{n,\kv}~\vec{A}_{nn}
\eec
in terms of the diagonal elements of the non-Abelian Berry connection ${\vec{A}_{nm} = i \bra{u_{n,\kv}}\ket{\nabla_\kv u_{m,\kv}}}$, integrated over the Brillouin zone (BZ), where $\ket{u_{n,\kv}}$ are the cell-periodic parts of the Bloch states. 
$e$ is an elementary charge and $f_{n,\kv}$ denotes a temperature-dependent occupation factor given by the Fermi-Dirac distribution. 
To describe shift currents, we define the `shift vectors'
\beq{eq:shift_vector}
\vec{R}_{mn}^{a} \equiv \vec{A}_{mm} - \vec{A}_{nn} - i\nabla_{\kv} \text{Arg}~ (A_{mn}^a)
\eec
which, if the last term is negligible and the $a$ index is omissible, can be thought of as the change in electronic polarization induced by a photoexcitation (see Methods). 
The shift photoconductivity $\s$ in response to linearly polarized light is given by~\cite{Sipe1993, Sipe2000, Ahn2021}
\beq{eq:shift}
\s^{c,aa}(\om) = -\frac{2\pi e^3}{\hbar^2} \sum_{m,n} \intBZ \frac{\dd^2 \kv}{(2\pi)^2} \d(\om - \om_{mn}) f_{nm,\textbf{k}} ~R^{c,a}_{mn} ~|A^{a}_{mn}|^2
\eec
where ${f_{nm,\kv} = f_{n,\kv} - f_{m,\kv}}$, $n$ ($m$) denotes a band with energy $E_{n,\kv}$ ($E_{m,\kv}$), and ${\omega_{mn} \equiv (E_{m,\kv} -  E_{n,\kv})/\hbar}$ are the frequencies of optical transitions. 
The shift photocurrent $\boldsymbol j$ is then given by~\cite{Sipe1993, Sipe2000}
\beq{}
j^{\, c} = 2 \s^{c,aa}(\om) \mathcal{E}^a(\om) \mathcal{E}^a(-\om)
\eec
where $\mathcal{E}^a(\om)$ are the components of an AC electric field of incident light with frequency $\om$ and Einstein summation convention is assumed.

It was recently proposed that shift currents can be used as an experimental diagnostic tool for quantum geometry in two-dimensional materials~\cite{Ma2023}, and can have nontrivial spatial dependence in supercells such as moir\'e superlattices, forming complex structures such as vortices~\cite{Chen2023}.
We propose that the vortices in the shift current are a direct result of the topological polarization textures that form in these materials: a~consequence of the interplay between polarization and shift currents~\cite{Fregoso2017, resta2024geo}.

\begin{figure*}[t!]
\includegraphics[width=\textwidth]{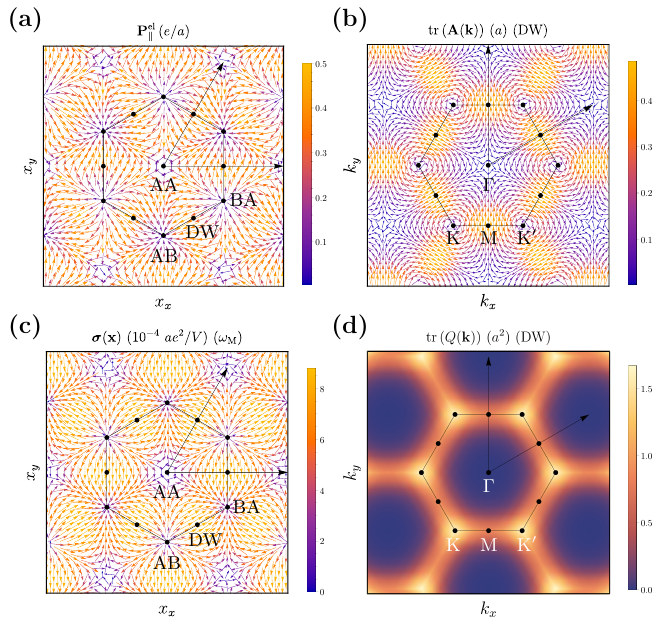}
\caption{
{\bf Polarization and shift photocurrent in twisted boron nitride.}
{\bf (a)} In-plane polarization of t-hBN contributed by electrons. The axes labels $x_x$ and $x_y$ are the components of the stacking vector $\xv$. The polarization texture was obtained using Eq.~\eqref{eq:KSV} with Berry connections calculated from the wavefunctions obtained by diagonalizing the four-band TB model (see Methods) and using parametrizations detailed in SI. {\bf (b)} Trace of the Berry connection in momentum-space over occupied bands, evaluated for the DW stacking.
The dominant contribution to the polarization occurs around the M points on the BZ edge.
{\bf (c)} Plot of the shift photoconductivity vector $\sv(\rv)$ in real space.
The photoconductivities are evaluated at a transition energy of $\om_{\rm M} = 6$ eV, and are antiparallel to the in-plane polarization.
The shift vortex features are stable for a range of frequencies around $\om_{\rm M}$ (see SI).
{\bf (d)} Trace of the QGT over spatial indices, summed over all interband transitions (for the DW stacking). The dominant QGT contributions arise from the regions with enhanced $\vec{A}_{nn}(\kv)$ (near M points), which fortifies the shift current-electric polarization correspondence near $\om_{\rm M}$.
}
    \label{Fig2}
\end{figure*}

In materials such as bilayer hBN or TMDs, twisting about the artificial rhombohedral (parallel) stacking results in a network of MPDs formed by the stacking domains \cite{li2017binary,bennett2022electrically,bennett2022theory}.
The supercell consists of four distinct regions: the non-polar AA stackings, which are energetically unstable but pinned by the geometry of the superlattice, the AB/BA domains, which are energetically favorable and have maximum out-of-plane polarization, and the domain walls (DWs)~\cite{Carr2020review, bennett2023polar}, which act as solitons separating the AB and BA domains.
These stackings are sketched in Fig.~\ref{Fig1} for bilayer hBN.
The MPDs also have an in-plane polarization texture, which is largest along the domain walls, see Fig.~\ref{Fig2} (a).
The dominant contribution of the Berry connection to the in-plane polarization occurs at the edge of the BZ, see Fig.~\ref{Fig2} (b).
Combining the in-plane and out-of-plane components, the polarization field exhibits topologically nontrivial winding, forming a network of merons and anti-merons, with winding numbers $Q=\pm\frac{1}{2}$~\cite{bennett2023polar, bennett2023theory}. The winding numbers $Q$ can be calculated by integrating the local winding of the normalized polarization~\cite{bennett2023theory} on a discretized grid (see SI), following the methodology described in Ref.~\cite{bennett2023polar}. 
The topological index, i.e. the wrapping number $Q$, is exactly (half-)quantized, as the local polarization cannot wind across a moir\'e domain in an arbitrary way, given the stacking symmetries of the moir\'e crystal and because of the periodicity of the supercells.

We observe that the shift photoconductivity vector field, defined as
\beq{eq:sigma-vector}
\sv(\rv) = 
\begin{bmatrix}
\sigma^{x,xx}(\rv) + \sigma^{x,yy}(\rv) \\
\sigma^{y,xx}(\rv) +  \sigma^{y,yy}(\rv)
\end{bmatrix}
\eec
directly coincides with the in-plane polarization field. 
Combining the responses to both $x$- and $y$-polarized electric fields allows for an analytical averaging of the shift vectors over all sections of the BZ where optical transitions are dominant (see SI for additional details). 
Since the \textit{averaged} shift vectors $\vec{R}_{mn}$ reflect the local electronic polarization, the components of $\sv(\rv)$ serve as a probe for the local polarization components, provided that the optical transition matrix elements are comparable in regions of the BZ that contribute most to the electronic polarization. 

\begin{figure*}[t]
\centering
\includegraphics[width=\linewidth]{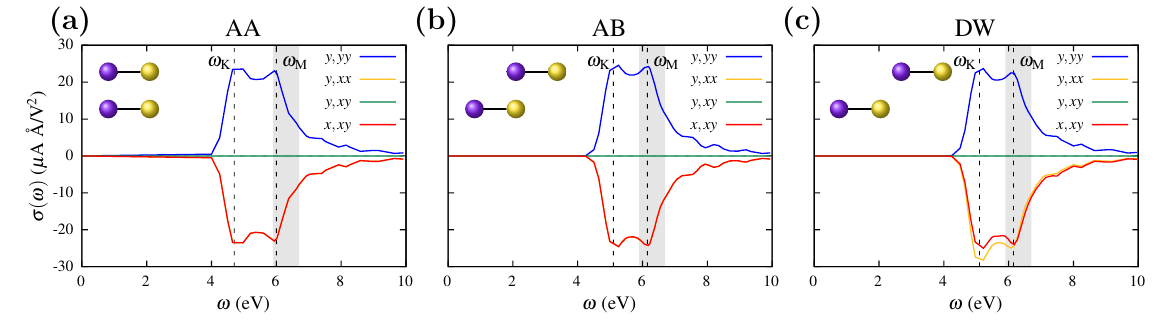}
\caption{
{\bf Spectrally resolved shift photoconductivity.}
Components of $\s$ defining the shift photoconductivity vector [Eq.~\eqref{eq:sigma-vector}]
as a function of light frequency $\omega$, at different relative stackings: 
{\bf (a)} AA, 
{\bf (b)} AB, and 
{\bf (c)} DW.
The $\s^{y,yy}$, $\s^{y,xx}$, $\s^{y,xy}$, and $\s^{x,xy}$ components are plotted in blue, yellow, green, and red respectively. 
These local shift photoconductivities have similar qualitative features for the different relative stackings.
The first peak at $\om_{\rm K} \approx 5~\text{eV}$ in all stackings arises from transitions at the K point and makes minor contributions to the electronic polarization. 
The second peak at $\om_{\rm M} \approx 6~\text{eV}$ corresponds to the photoexcitations of the electrons near the M point and contributes the dominant part of the electronic polarization, as reflected by their shift vectors $\vec{R}_{mn}$.
The range of frequencies, for which the shift current vortices due to the electronic polarization textures occur, is highlighted in gray. For AA and AB stackings, the $\sigma^{y,xx}$ and $\sigma^{x,xy}$ components are equivalent by symmetry (see SI). 
}
\label{Fig3}
\end{figure*}

We illustrate the correspondence between polarization textures and shift photoconductivity textures using a tight-binding (TB) model of twisted bilayer hBN (t-hBN).
The model consists of four bands, representing the two valence (conduction) bands closest to the Fermi level, of $2p_z$ character on the N (B) atoms in each layer (see Methods).
The local polarization and shift conductivity are calculated using the configuration space method, under the approximation that for small twist angles (large supercells), the local stacking order changes slowly and smoothly, and local environments are well described by a relative shift between two commensurate layers \cite{carr2018relaxation,bennett2023polar, bennett2023theory}.
The shift photoconductivity vector is shown in Fig.~\ref{Fig2} (c), at a frequency of $\om_{\rm M} = 6$ eV, roughly corresponding to the resonant transitions at the M point of the BZ.
We note that
$\sv(\rv)$ is exactly antiparallel to $\Pv(\rv)$:
the current flows out of (into) the AB (BA) domains, has largest magnitude along the domain walls, and forms vortices around the AA stacking regions. 
This correspondence is most strongly observed within a range of light frequencies where the resonant transitions occur in regions of the BZ which contribute most dominantly to the in-plane polarization, see Fig.~\ref{Fig2} (b).

The origin of this correspondence can be traced to the interplay between the interband transition rates described by the QGT, $Q^{ab}_{mn} = A^{a}_{nm} A^{b}_{mn}$ (see Methods), and the non-Abelian Berry connection $\vec{A}_{nm}(\kv)$.
The dominant optical transition rates occur at the M points (edges) and the K/K$^\prime$ points (corners) of the BZ (K and K$^\prime$ correspond to different valleys) as elucidated by the spatial trace of the QGT in Fig.~\ref{Fig2} (d) for the DW stacking. 
The shift photocurrents combine both the QGT and the diagonal elements of the non-Abelian Berry connection $\vec{A}_{nm}(\kv)$. 
However, as shown in Fig.~\ref{Fig2}~(b),
$\vec{A}_{nn}(\kv)$ forms a vortex at the K point. 
As its magnitude smoothly goes to zero at the vortex center, transitions at the K point contribute less to both the electronic polarization and the shift photocurrents.
In contrast, at the M points, which are the saddle points in the effective band structures and the trace of the QGT, $\vec{A}_{nn}(\kv)$ flows with largest magnitude [see Fig.~\ref{Fig2} (b)], contributing most strongly to the electronic polarization and shift photoconductivity. 
The correspondence between polarization and shift photoconductivity occurs in a range of frequencies near $\om_{\rm M}$, wherein both the QGT and the diagonal elements of the non-Abelian Berry connection are significant, resulting in large photovoltaic shift responses. 
Since the same regions also contribute strongly to the electronic polarization that is given by the Berry connection in Eq.~\eqref{eq:KSV}, these shift currents arise directly from the electronic polarization and contribute to the aforementioned correspondence revealed in Figs.~\ref{Fig2} (a) and (c) (see SI for details).

The spectral dependence of $\s$ is shown in Fig.~\ref{Fig3}.
The spectrally-resolved components $\s^{c,ab}$ that determine the shift photoconductivity vector in Eq.~\eqref{eq:sigma-vector} are shown for the AA, AB and DW stackings. 
The elements $\sigma^{x,xy}$ and $\sigma^{y,xy}$ encode the responses to both linearly and circularly polarized light, in their real and imaginary parts, respectively~\cite{Ahn2020}.
Since electric polarization is specifically related to linear light polarization, we focus only on calculating the real parts of the associated shift photoconductivities.

\begin{figure*}[t!]
\centering
\includegraphics[width=\linewidth]{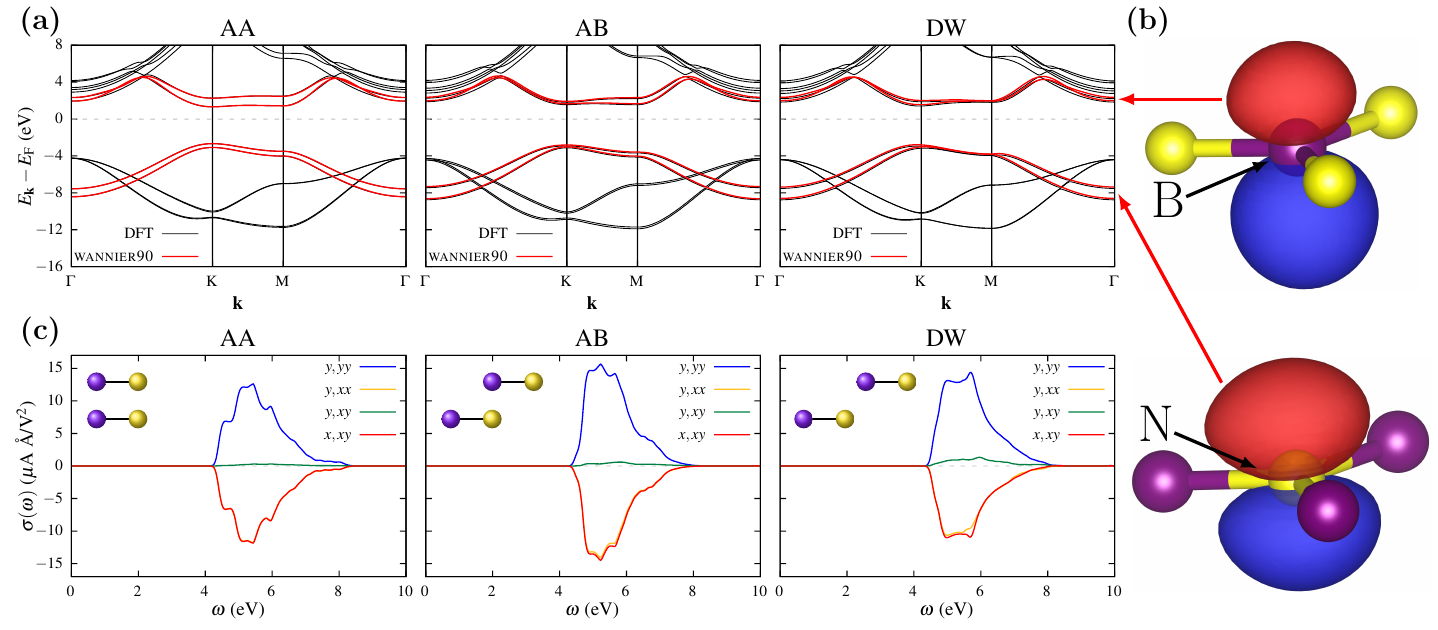}
\caption{\textbf{First-principles calculations of shift photoconductivity.}
{\bf (a)} Electronic band structure of bilayer hBN for the AA, AB and DW stackings. 
The bands obtained from first-principles calculations are shown in black.
The Wannierized bands of the $p_z$ orbitals are shown in red.
{\bf (b)} Illustration of the real-space Wannier functions corresponding to the $p_z$ orbitals on the B (purple) and N (yellow) atoms, which yield the conduction and valence Wannier bands, respectively. 
{\bf (c)} Frequency-resolved shift currents obtained from the Wannier functions for the AA, AB and DW stackings. The $\s^{y,yy}$, $\s^{y,xx}$, $\s^{y,xy}$, and $\s^{x,xy}$ components are plotted in blue, yellow, green, and red respectively.
}
\label{Fig4}
\end{figure*}

The first peak occurs at $\om_{\rm K} \approx 5$ eV for all three stackings, which corresponds to transitions at the K point.
While this peak represents a relatively significant contribution to the shift response, as mentioned previously, it does not reflect the electronic polarization, as the Berry connection forms vortices around the K points.
The most significant contributions to $\Pv$ come from the peak at $\om_{\rm M}$ which arises due to transitions at the M point.
The M point corresponds to a van Hove singularity in the joint density of states (JDOS), where the Berry connection is also the largest, and thus makes the strongest contribution to the electronic polarization (see SI). 
As noted earlier, a range of frequencies near $\om_{\rm M}$ achieves the desired shift current vortices for the electronic polarization correspondence and is highlighted in Fig.~\ref{Fig3}.
In Fig.~\ref{Fig3} we also observe additional peaks at higher frequencies.

The results in Figs.~\ref{Fig2} and \ref{Fig3} were validated with first-principles calculations of bilayer hBN as a function of relative stacking (see Methods).
The electronic structure contains two valence bands and two conduction bands near the Fermi level, arising from the $2p_z$ orbitals of the N and B atoms in the unit cell, respectively (see Fig.~\ref{Fig4}).
We obtain the maximally localized Wannier functions (MLWFs)~\cite{marzari2012maximally, pizzi2020wannier90} of these 4 states by projecting the Bloch states onto $p_z$ orbitals centered on each atom and numerically minimizing the spread of the Wannier functions.
The shift current was then calculated using Wannier interpolation~\cite{Ibanez2018}.
The spectrally resolved shift photoconductivities are shown in Fig.~\ref{Fig4} for the AA, AB and DW stackings, in good agreement with Fig.~\ref{Fig3}.

\section{Discussion}
In this work, we show that topological polarization textures can result in exotic nonlinear optical responses, namely a shift photoconductivity with a vortex-like pattern.
We illustrate this concept using bilayer hBN, the prototypical vdW ferroelectric, as an example, although the results can be generalized to other materials.
The general shape of the stacking-dependent polarization can be determined solely from symmetry analysis of the space groups of the different stackings \cite{ji2023general,bennett2023polar}:
by symmetry the shape of the polarization textures in TMDs twisted about the rhombohedral stacking is identical to the textures in t-hBN.
Based on the generality of the presented theoretical framework, we expect any twisted bilayer with topological polarization to exhibit shift photoconductivity vortices, such as TMDs~\cite{vu2024imaging}. 
Experimentally, these could be retrieved from the optical frequency windows determined by the band gaps of TMDs (1-2~eV), which are practically more accessible than the band gaps of the twisted hBN (5-6~eV) studied here.
While the large band gap of hBN may pose challenges for the experimental realization of these shift photoconductivity vortices, by selecting different materials the gap can effectively be tuned to suit the experimental setup.

In addition to the polarization, the shift current texture is also constrained by symmetry.
As shown in Figs.~\ref{Fig3} and \ref{Fig4}, the $\sigma^{y,xy}$ component, which describes the coupling of both $x$ and $y$ components to in-plane electric fields, is vanishingly small for all stackings.
Furthermore, the $\Ct$ rotation symmetry of the lattice constrains several components of the shift photoconductivity tensor: $\s^{x,xx} = -\s^{x,yy} = -\s^{y,xx} = -\s^{y,xy}$ \cite{Chen2023}. 
Moreover, $\sigma^{y,yy} = -\sigma^{x,xy}$, as demonstrated in Fig.~\ref{Fig3}.
While our results satisfy these symmetry-imposed constraints, our model relies on the configuration space approximation, which explicitly assumes the locality of the shift photocurrents and the polarization within the supercell~\cite{bennett2023polar}.
However, the configuration space approximation yields polarization textures in excellent agreement with experiment~\cite{ko2023operando}, and our calculations of the shift photoconductivities are in excellent agreement with large-scale calculations of twisted bilayers~\cite{Chen2023}.

We show that experimentally measurable shift photocurrents can be used to directly map out the in-plane polarization component of MPDs at characteristic light frequencies. The local correspondence between polarization and shift current proposed here may be used to optically probe topological polarization textures with sub-diffraction photocurrent spectroscopy techniques. In the previous studies, such polarization textures have proven difficult to observe experimentally~\cite{vu2024imaging}.
The photocurrent microscopy techniques provide a tunable modern platform for spatially resolving photoexcited quantities on the nanometer scale~\cite{lang1978scanning, rauhut2012antenna, Hesp_2021, Ma2023, Xiang2024},
which has recently been successfully applied to spatially resolve the photocurrents in twisted \chem{WS_2}~\cite{li2024imaging}.
Although direct optical experiments are limited to lengthscales of order 100~nm, sub-diffraction photocurrent spectroscopy techniques can resolve details on lengthscales of order 10~nm (with an approximate resolution of 7.5~nm)~\cite{Hesp_2021}, and should be capable of resolving the photocurrent vortex structures predicted in this work.
Resolving the flow of the shift photoconductivity would indirectly signal the in-plane polarization and the topological nature of the MPDs in moir\'e materials.
This technique could also be used to probe topological polarization textures in other materials, such as oxide perovskites~\cite{junquera2023topologicaly, sanchez20242d}.

The shift photoconductivities arising from in-plane electronic polarization in this work include only the photoexcitation part to the shift currents~\cite{Belinicher1982, zhu2024anomalousshiftopticalvorticity}.
We note that such photoexcitation contributions can be probed in transient responses with sub-picosecond resolution~\cite{Sotome2019}.
In other scenarios, namely over longer timescales, there are additional contributions from phonon and impurity-dependent intraband scattering, as well as from carrier relaxation~\cite{Belinicher1982, zhu2024anomalousshiftopticalvorticity}. 
Finally, we find that the injection currents, i.e.~second-order photocurrents arising from photoinduced changes of group velocities~\cite{Sipe1993, Sipe2000}, are negligible in response to linearly polarized light (see SI). 
This further highlights the feasibility of experimentally probing electronic polarization textures using the correspondence to the shift photoconductivities.

In summary, we show that there is a correspondence between shift current vortices and topological polarization textures. 
We propose that second-order bulk shift photocurrents can be used to deduce the presence of in-plane electronic  polarization in 2D ferroelectrics such as t-hBN, facilitating the experimental observation of topological polarization in marginally-twisted bilayers. 
We anticipate that this correspondence will therefore play a key role in uncovering the new landscape of polar domains as a platform for novel physical effects.


\let\section\SectionOriginal



\section*{Methods}

\subsection*{Quantum-geometric relations}
The quantum geometric details associated with the correspondence between the shift current and the electronic contribution to material's electronic polarization are summarized in this section. 
The non-Abelian (multiband) Berry connection $A^a_{nm}$~\cite{vanderbilt2018berry} is obtained within the TB model. The polarization can then be related as a volume integral of the trace of the non-Abelian Berry connection over occupied (`occ') states, which in turn, on projecting on the direction $\beta$, can be written in terms of Berry phases: $P_\beta = -\frac{e}{a_\beta} \sum_{k_\perp} \phi(k_\perp)$. The Berry phases are defined in terms of the Berry connection as,
\beq{}
    \phi(k_\perp) = \frac{1}{2\pi} \int \dd k_\beta \sum^{\text{occ}}_n A^\beta_{nn} = \frac{1}{2\pi} \int \dd k_\beta~\text{tr}~A^{\beta},
\eeq
where ${a_\beta}$ is the unit cell size in the direction $\beta$. We note that the above are gauge-invariant objects. The non-Abelian Berry connection defines the Hermitian connection associated with the shift photoconductivities~\cite{Ahn2021},
\beq{}
    C^{abc}_{nm} = A^a_{nm} \mathcal{D}_{b} A^c_{mn},
\eeq
where the covariant derivative is defined with diagonal elements of the Berry connection as $\mathcal{D}_{a} = \partial_{k_a} - i(A^a_{mm}-A^a_{nn})$. In terms of Hermitian connections $C^{abc}_{nm}$, the shift photoconductivities can be written for a two-dimensional system as~\cite{Ahn2021}
\beq{}
    \sigma^{c,ab} (\om) = -\frac{ie^3}{8 \pi \hbar^2} \sum_{m,n} \int \dd^2 \textbf{k}~\delta(\om - \om_{mn}) f_{nm,\kv} [C_{nm}^{acb} - (C_{nm}^{bca})^* ].
\eeq
The above reduces to the shift vector formula [Eq.~\eqref{eq:shift}], upon recognizing that the shift vectors read componentwise as
\beq{}
    R^{c,a}_{mn} = A^c_{mm}-A^c_{nn} - i \partial_{k_c} \text{Arg}~ (A^a_{mn}).
\eeq
In the systems central to this work, under an appropriate gauge, and for $a=b$, the last term can be neglected, as we also detail and numerically demonstrate in the SI. In particular, the contribution due to the last term vanishes identically in an optical gauge~\cite{Fregoso2017} assuming topologically trivial bands. This then allows the reduction of the shift vector to an entity with a single spatial index: ${R^{c}_{mn} \approx A^c_{mm}-A^c_{nn}}$.

Beyond the shift vector, the shift photoconductivity notably involves the quantum-geometric tensor (QGT) $Q^{ab}_{mn}$ defined in terms of non-Abelian Berry connection elements as 
\beq{}
    Q^{ab}_{mn} = A^{a}_{nm} A^{b}_{mn},
\eeq
which captures the interband transition rates. For more details on QGT and its relations to optics and geometry, see SI.

\subsection*{Tight-binding model}
Following Refs.~\cite{Yu2023} and \cite{jankowski2024polarization}, we construct an effective tight-binding model to describe twisted moiré hBN (t-hBN) bilayer. To that end, we note that the low-energy Hamiltonian,  describing the $p_z$-orbital bands below and above the Fermi level can be written as
\begin{equation} \label{eqn:big_TB_H}
    H =
\begin{pmatrix}
    \frac{\Delta}{2} & t_\kv & t_{B B, \kv} & t_{B N, \kv} \\
    t^*_\kv & -\frac{\Delta}{2} & t_{N B, \kv}  & t_{N N, \kv}\\
    t^*_{B B, \kv} & t^*_{N B, \kv} & \frac{\Delta}{2} & t_\kv\\
    t^*_{B N, \kv} &  t^*_{N N, \kv} & t^*_\kv & -\frac{\Delta}{2}\\
\end{pmatrix},
\end{equation}
where we assumed a basis of cell-periodic Bloch states of the top (t) and bottom (b) layer boron and nitrogen atoms: ${ \ket{B_t}, \ket{N_t}, \ket{B_b}, \ket{N_b}}$. In the above, the implicit layer indices were dropped for simplicity. 

In a moir\'e insulator with well-preserved gaps, such as {t-hBN}, the off-diagonal $2 \times 2$ blocks can be treated as perturbations. 
The monolayer problem can be solved first (setting the $2\times 2$ off-diagonal blocks as zero) to obtain unperturbed eigenstates $\ket{u^{t/b}_{c/v,\kv}}$~\cite{Yu2023, jankowski2024polarization}. 
As a next step, the interlayer couplings can be included within the configuration space approximation as perturbations~\cite{bennett2023polar}. Correspondingly, one obtains perturbed eigenstates $\ket{\Tilde{u}^{t/b}_{c/v,\kv}}$, from the perturbation theory in the interlayer tunnelling constituting the off-diagonal terms. 
Such a transformation from the unperturbed to the perturbed eigenstates is an SU(4) transformation and can be described by the matrix M such that $\ket{\tilde{u}_{\kv}} = M^T \ket{u_{\kv}}$. Here, the vectors $\ket{\tilde{u}_{\kv}}$ and $\ket{u_{\kv}}$ include the eigenstates in conduction and valence bands $(c, v)$, with the top and bottom $(t, b)$ layer flavours. 
Second-order perturbation theory then dictates that $M$ is given by 
\begin{equation} \label{eqn:SU(4)_M}
    M = 
\begin{pmatrix}
    1 - \frac{1}{2} \Big| \frac{t^{bt}_{vc,\kv}}{\Delta_\kv} \Big|^2 & 0 & 0 & -\frac{t^{bt}_{vc,\kv}}{\Delta_\kv} \\
    0 &  1 - \frac{1}{2} \Big| \frac{t^{tb}_{vc,\kv}}{\Delta_\kv} \Big|^2  & \frac{(t^{tb}_{vc,\kv})^*}{\Delta_\kv} & 0 \\
    0 & -\frac{t^{tb}_{vc,\kv}}{\Delta_\kv} &  1 - \frac{1}{2} \Big| \frac{t^{tb}_{vc,\kv}}{\Delta_\kv} \Big|^2  & 0 \\
    \frac{(t^{bt}_{vc,\kv})^*}{\Delta_\kv} & 0 & 0 &  1 - \frac{1}{2} \Big| \frac{t^{bt}_{vc,\kv}}{\Delta_\kv} \Big|^2 
\end{pmatrix},
\end{equation}
where $\Delta_\kv$ is a local energy gap and the interlayer coupling constants $t^{tb}_{vc,\kv}(\vec{x})$,  $t^{bt}_{vc,\kv}(\vec{x})$ hybridize the hopping terms of the Hamiltonian $H$ in Eq.~\eqref{eqn:big_TB_H} (${t_{B N, \kv}, t_{N B, \kv}, t_{B B, \kv}, t_{N N, \kv}}$). 
The hybridized hoppings are given by the stacking $\vec{x}$ and the $\textbf{k}$-dependent factors, as detailed explicitly in SI.

The above description ensures that any dependence on stacking is entirely encapsulated in the SU(4) transformation $M$, and all stacking-dependent properties can therefore be rewritten as functions of $M$. 

Since all quantum-geometric quantities and subsequent optical properties introduced in the further section, depend on the non-Abelian Berry connection, it is useful to rewrite it in terms of the unperturbed connection $\textbf{A}$ and the SU(4) transformation M as 
\beq{}
    \tilde{\textbf{A}} = -i \, \, M^{\dagger} \nabla_{\kv} M + M^{\dagger} \textbf{A} M.
\eeq
All stacking-dependent electronic polarizations $\textbf{P}(\textbf{x})$ and shift photoconductivities $\sigma^{c,ab}(\textbf{x})$ with $a,b,c = x,y$, can be extracted from the quantum-geometric relations encoded by the SU(4)-modified non-Abelian Berry connection matrix $\tilde{\textbf{A}}$. To study the quantum-geometric relations, we replace the unperturbed connection with the modified one, by relabelling: $\tilde{\textbf{A}} \rightarrow \textbf{A}$.

The tight-binding calculations were carried out assuming the form of the real-space hoppings $t_{XY}$, with $X,Y$ denoting the B, N atoms, to be an exponential decay in $|\xv|$ with an upper cutoff, following Refs.~\cite{jankowski2024polarization} and \cite{Yu2023}. 
In order to obtain a faithful description of hBN with an interlayer spacing of $0.33 \, \text{nm}$, the tight-binding parameters have been chosen following Ref.~\cite{Yu2023} as $\Delta =4.5 \, \text{eV}$, $t = 2.0 \, \text{eV}$, for the intralayer nearest-neighbor hoppings, and $t^0_{BN} = t^0_{NB} = 1.28 \, \text{eV}$, $t^0_{BB} = 0.8 \, \text{eV}$, and $t^0_{NN} = 0.6 \, \text{eV}$ for the interlayer hopping parameters. 
The latter parameters were further regularized to account for the relative displacements associated with the local stackings. 
The interlayer coupling regularizations have been chosen following Ref.~\cite{jankowski2024polarization}. 
In addition to the chosen parameters, the Fermi-Dirac occupations are regarded as $f_{v \kv} = 1$ for the valence bands and $f_{c \kv} = 0$ for the conduction bands, since the energy gap between the conduction and valence band $\Delta_\kv \approx 4.5 \, \text{eV} \gg k_B T$ for temperatures $T \lesssim 10^4 \, \text{K}$. 

\subsection*{First-principles calculations}

First-principles density functional theory (DFT) calculations were performed to simulate bilayer hBN, in the rhombohedral (aligned) stacking, using the {\sc abinit} \cite{gonze2016,gonze2020} code.
Norm-conserving \cite{norm_conserving} {\sc psml} \cite{psml} pseudopotentials were used, obtained from Pseudo-Dojo \cite{pseudodojo}. 
{\sc abinit} employs a plane-wave basis set, which was determined using a kinetic energy cutoff of $1000$ eV. 
A Monkhorst-Pack $\kv$-point grid \cite{mp} of $12 \times 12 \times 1$ was used to sample the Brillouin zone. 
The revPBE exchange-correlation functional was used \cite{zhang1998comment}, and the vdw-DFT-D3(BJ) \cite{becke2006simple} correction was used to treat the vdW interactions between the layers.

In order to sample the relative stackings between the layers in `configuration space' \cite{carr2018relaxation},
the top layer was translated along the unit cell diagonal over the bottom layer, which was held fixed. 
The relative stackings were sampled in 2D using a grid of $6\times 6$, which explicitly includes the high symmetry stackings: the AA stacking, where the two layers are perfectly aligned, the AB and BA where the opposite atoms in neighboring layers are vertically aligned, given by a relative shift of $x=\frac{1}{3}$ or $\frac{2}{3}$ of a unit cell diagonal, respectively, and domain wall (DW) stacking, given by a shift of $x=\frac{1}{2}$ of a unit cell diagonal.
At each point a geometry relaxation was performed to obtain the equilibrium layer separation, while keeping the in-plane atomic positions fixed. 

Maximally localized Wannier functions (MLWF) were then constructed for the two valence bands and two conduction bands closest to the Fermi level, using the interface between {\sc abinit} and {\sc Wannier90}~\cite{marzari2012maximally, pizzi2020wannier90}.
The valence (conduction) bands are of N (B) $2p_z$ character. (see SI).
The initial projections were made onto the $2p_z$ orbitals of the four atoms in the bilayer unit cell.
A disentanglement procedure was performed to obtain MLWFs for the entangled bands near the Fermi level, using a frozen energy window which contains only the four bands closest to the Fermi level, and an outer energy window which contains those bands everywhere in the BZ.
After the disentanglement procedure, the spread of the Wannier functions was then minimized.
The shift current was then obtained using the Wannier interpolation~\cite{Ibanez2018}.
Calculations were repeated to obtain the MLWFs and shift currents as a function of relative stacking between the layers.

\bigskip

\section*{Data Availability}
The datasets generated and analyzed in this study are available from the corresponding author upon request.

\section*{Code Availability}
The data presented in this study were generated using theoretical models as well as free and open-source first-principles packages as described in the Methods section. 

\section*{Acknowledgments}
The authors acknowledge G.~Chaudhary and A.~Mishchenko for helpful discussions. 
A.~A.~acknowledges funding from the Cambridge International Scholarship awarded by the Cambridge Trust.
W.~J.~J.~acknowledges funding from the Rod Smallwood Studentship at Trinity College, Cambridge. 
D.~B.~acknowledges the US Army Research Office (ARO) MURI project under grant No.~W911NF-21-0147 and the Simons Foundation award No.~896626. 
R.-J.~S.~acknowledges funding from a New Investigator Award, EPSRC grant EP/W00187X/1. 
\text{R.-J.~S.}~also acknowledges funding from a EPSRC ERC underwrite grant EP/X025829/1, as well as Trinity College, Cambridge.

\section*{Author Contributions}
A.~A.~performed the tight-binding calculations with input from W.~J.~J.~and R.-J.~S. A.~A., W.~J.~J.~and R.-J.~S.~developed the theory analysis of the local shift currents and relation to quantum geometry.
D.~B.~performed the first-principles calculations.
All authors contributed to the analysis and interpretation of the results and to the writing of the paper.

\section*{Competing Interests}
The authors declare no competing interests.

\bibliographystyle{naturemag}
\bibliography{references.bib}

\begin{thebibliography}{10}
\expandafter\ifx\csname url\endcsname\relax
  \def\url#1{\texttt{#1}}\fi
\expandafter\ifx\csname urlprefix\endcsname\relax\def\urlprefix{URL }\fi
\providecommand{\bibinfo}[2]{#2}
\providecommand{\eprint}[2][]{\url{#2}}

\bibitem{Bistritzer2011}
\bibinfo{author}{Bistritzer, R.} \& \bibinfo{author}{MacDonald, A.~H.}
\newblock \bibinfo{title}{Moire bands in twisted double-layer graphene}.
\newblock \emph{\bibinfo{journal}{PNAS}} \textbf{\bibinfo{volume}{108}}, \bibinfo{pages}{12233} (\bibinfo{year}{2011}).
\newblock \urlprefix\url{https://www.pnas.org/doi/10.1073/pnas.1108174108}.

\bibitem{carr2017twistronics}
\bibinfo{author}{Carr, S.} \emph{et~al.}
\newblock \bibinfo{title}{Twistronics: manipulating the electronic properties of two-dimensional layered structures through their twist angle}.
\newblock \emph{\bibinfo{journal}{Phys. Rev. B}} \textbf{\bibinfo{volume}{95}}, \bibinfo{pages}{075420} (\bibinfo{year}{2017}).
\newblock \urlprefix\url{https://doi.org/10.1103/PhysRevB.95.075420}.

\bibitem{cao2018unconventional}
\bibinfo{author}{Cao, Y.} \emph{et~al.}
\newblock \bibinfo{title}{Unconventional superconductivity in magic-angle graphene superlattices}.
\newblock \emph{\bibinfo{journal}{Nature}} \textbf{\bibinfo{volume}{556}}, \bibinfo{pages}{43} (\bibinfo{year}{2018}).
\newblock \urlprefix\url{https://doi.org/10.1038/nature26160}.

\bibitem{nuckolls2020strongly}
\bibinfo{author}{Nuckolls, K.~P.} \emph{et~al.}
\newblock \bibinfo{title}{Strongly correlated {C}hern insulators in magic-angle twisted bilayer graphene}.
\newblock \emph{\bibinfo{journal}{Nature}} \textbf{\bibinfo{volume}{588}}, \bibinfo{pages}{610--615} (\bibinfo{year}{2020}).
\newblock \urlprefix\url{https://doi.org/10.1038/s41586-020-3028-8}.

\bibitem{wu2021chern}
\bibinfo{author}{Wu, S.}, \bibinfo{author}{Zhang, Z.}, \bibinfo{author}{Watanabe, K.}, \bibinfo{author}{Taniguchi, T.} \& \bibinfo{author}{Andrei, E.~Y.}
\newblock \bibinfo{title}{{C}hern insulators, van hove singularities and topological flat bands in magic-angle twisted bilayer graphene}.
\newblock \emph{\bibinfo{journal}{Nat. Mater.}} \textbf{\bibinfo{volume}{20}}, \bibinfo{pages}{488--494} (\bibinfo{year}{2021}).
\newblock \urlprefix\url{https://doi.org/10.1038/s41563-020-00911-2}.

\bibitem{xie2021fractional}
\bibinfo{author}{Xie, Y.} \emph{et~al.}
\newblock \bibinfo{title}{Fractional {C}hern insulators in magic-angle twisted bilayer graphene}.
\newblock \emph{\bibinfo{journal}{Nature}} \textbf{\bibinfo{volume}{600}}, \bibinfo{pages}{439--443} (\bibinfo{year}{2021}).
\newblock \urlprefix\url{https://doi.org/10.1038/s41586-021-04002-3}.

\bibitem{zeng2023thermodynamic}
\bibinfo{author}{Zeng, Y.} \emph{et~al.}
\newblock \bibinfo{title}{Thermodynamic evidence of fractional {C}hern insulator in moir{\'e} {M}o{T}e$_2$}.
\newblock \emph{\bibinfo{journal}{Nature}} \textbf{\bibinfo{volume}{622}}, \bibinfo{pages}{69--73} (\bibinfo{year}{2023}).
\newblock \urlprefix\url{https://doi.org/10.1038/s41586-023-06452-3}.

\bibitem{park2023observation}
\bibinfo{author}{Park, H.} \emph{et~al.}
\newblock \bibinfo{title}{Observation of fractionally quantized anomalous {H}all effect}.
\newblock \emph{\bibinfo{journal}{Nature}} \textbf{\bibinfo{volume}{622}}, \bibinfo{pages}{74--79} (\bibinfo{year}{2023}).
\newblock \urlprefix\url{https://doi.org/10.1038/s41586-023-06536-0}.

\bibitem{tong2018skyrmions}
\bibinfo{author}{Tong, Q.}, \bibinfo{author}{Liu, F.}, \bibinfo{author}{Xiao, J.} \& \bibinfo{author}{Yao, W.}
\newblock \bibinfo{title}{Skyrmions in the moir{\'e} of van der {W}aals 2{D} magnets}.
\newblock \emph{\bibinfo{journal}{Nano Lett.}} \textbf{\bibinfo{volume}{18}}, \bibinfo{pages}{7194--7199} (\bibinfo{year}{2018}).
\newblock \urlprefix\url{https://doi.org/10.1021/acs.nanolett.8b03315}.

\bibitem{hejazi2020noncollinear}
\bibinfo{author}{Hejazi, K.}, \bibinfo{author}{Luo, Z.-X.} \& \bibinfo{author}{Balents, L.}
\newblock \bibinfo{title}{Noncollinear phases in moir{\'e} magnets}.
\newblock \emph{\bibinfo{journal}{PNAS}} \textbf{\bibinfo{volume}{117}}, \bibinfo{pages}{10721--10726} (\bibinfo{year}{2020}).
\newblock \urlprefix\url{https://doi.org/10.1073/pnas.2000347117}.

\bibitem{song2021direct}
\bibinfo{author}{Song, T.} \emph{et~al.}
\newblock \bibinfo{title}{Direct visualization of magnetic domains and moir{\'e} magnetism in twisted 2{D} magnets}.
\newblock \emph{\bibinfo{journal}{Science}} \textbf{\bibinfo{volume}{374}}, \bibinfo{pages}{1140--1144} (\bibinfo{year}{2021}).
\newblock \urlprefix\url{https://doi.org/10.1126/science.abj7478}.

\bibitem{bennett2024stacking}
\bibinfo{author}{Bennett, D.} \emph{et~al.}
\newblock \bibinfo{title}{Stacking-engineered ferroelectricity and multiferroic order in van der waals magnets}.
\newblock \emph{\bibinfo{journal}{Phys. Rev. Lett.}} \textbf{\bibinfo{volume}{133}}, \bibinfo{pages}{246703} (\bibinfo{year}{2024}).
\newblock \urlprefix\url{https://link.aps.org/doi/10.1103/PhysRevLett.133.246703}.

\bibitem{stern2020interfacial}
\bibinfo{author}{Stern, M.~V.} \emph{et~al.}
\newblock \bibinfo{title}{Interfacial ferroelectricity by van der {W}aals sliding}.
\newblock \emph{\bibinfo{journal}{Science}} \textbf{\bibinfo{volume}{372}}, \bibinfo{pages}{1462} (\bibinfo{year}{2021}).
\newblock \urlprefix\url{https://www.science.org/doi/10.1126/science.abe8177}.

\bibitem{yasuda2021stacking}
\bibinfo{author}{Yasuda, K.}, \bibinfo{author}{Wang, X.}, \bibinfo{author}{Watanabe, K.}, \bibinfo{author}{Taniguchi, T.} \& \bibinfo{author}{Jarillo-Herrero, P.}
\newblock \bibinfo{title}{Stacking-engineered ferroelectricity in bilayer boron nitride}.
\newblock \emph{\bibinfo{journal}{Science}} \textbf{\bibinfo{volume}{372}}, \bibinfo{pages}{1458} (\bibinfo{year}{2021}).
\newblock \urlprefix\url{https://doi.org/10.1126/science.abd3230}.

\bibitem{wang2022interfacial}
\bibinfo{author}{Wang, X.} \emph{et~al.}
\newblock \bibinfo{title}{Interfacial ferroelectricity in rhombohedral-stacked bilayer transition metal dichalcogenides}.
\newblock \emph{\bibinfo{journal}{Nat. Nanotechnol.}} \textbf{\bibinfo{volume}{17}}, \bibinfo{pages}{367--371} (\bibinfo{year}{2022}).
\newblock \urlprefix\url{https://doi.org/10.1038/s41565-021-01059-z}.

\bibitem{weston2022interfacial}
\bibinfo{author}{Weston, A.} \emph{et~al.}
\newblock \bibinfo{title}{Interfacial ferroelectricity in marginally twisted 2{D} semiconductors}.
\newblock \emph{\bibinfo{journal}{Nat. Nanotechnol.}} \textbf{\bibinfo{volume}{17}}, \bibinfo{pages}{390--395} (\bibinfo{year}{2022}).
\newblock \urlprefix\url{https://doi.org/10.1038/s41565-022-01072-w}.

\bibitem{ko2023operando}
\bibinfo{author}{Ko, K.} \emph{et~al.}
\newblock \bibinfo{title}{Operando electron microscopy investigation of polar domain dynamics in twisted van der {W}aals homobilayers}.
\newblock \emph{\bibinfo{journal}{Nat. Mater.}} \bibinfo{pages}{1--7} (\bibinfo{year}{2023}).
\newblock \urlprefix\url{https://doi.org/10.1038/s41563-023-01595-0}.

\bibitem{molino2023ferroelectric}
\bibinfo{author}{Molino, L.} \emph{et~al.}
\newblock \bibinfo{title}{Ferroelectric switching at symmetry-broken interfaces by local control of dislocations networks}.
\newblock \emph{\bibinfo{journal}{Adv. Mater.}} \textbf{\bibinfo{volume}{35}}, \bibinfo{pages}{2207816} (\bibinfo{year}{2023}).
\newblock \urlprefix\url{https://doi.org/10.1002/adma.202207816}.

\bibitem{van2024engineering}
\bibinfo{author}{Van~Winkle, M.} \emph{et~al.}
\newblock \bibinfo{title}{Engineering interfacial polarization switching in van der {W}aals multilayers}.
\newblock \emph{\bibinfo{journal}{Nat. Nanotechnol.}} \bibinfo{pages}{1--7} (\bibinfo{year}{2024}).
\newblock \urlprefix\url{https://doi.org/10.1038/s41565-024-01642-0}.

\bibitem{yasuda2024ultrafast}
\bibinfo{author}{Yasuda, K.} \emph{et~al.}
\newblock \bibinfo{title}{Ultrafast high-endurance memory based on sliding ferroelectrics}.
\newblock \emph{\bibinfo{journal}{Science}} \bibinfo{pages}{eadp3575} (\bibinfo{year}{2024}).
\newblock \urlprefix\url{https://doi.org/10.1126/science.adp3575}.

\bibitem{bian2024developing}
\bibinfo{author}{Bian, R.} \emph{et~al.}
\newblock \bibinfo{title}{Developing fatigue-resistant ferroelectrics using interlayer sliding switching}.
\newblock \emph{\bibinfo{journal}{Science}} \bibinfo{pages}{eado1744} (\bibinfo{year}{2024}).
\newblock \urlprefix\url{https://doi.org/10.1126/science.ado1744}.

\bibitem{li2017binary}
\bibinfo{author}{Li, L.} \& \bibinfo{author}{Wu, M.}
\newblock \bibinfo{title}{Binary compound bilayer and multilayer with vertical polarizations: Two-dimensional ferroelectrics, multiferroics, and nanogenerators}.
\newblock \emph{\bibinfo{journal}{ACS Nano}} \textbf{\bibinfo{volume}{11}}, \bibinfo{pages}{6382--6388} (\bibinfo{year}{2017}).
\newblock \urlprefix\url{https://pubs.acs.org/doi/abs/10.1021/acsnano.7b02756}.

\bibitem{bennett2022electrically}
\bibinfo{author}{Bennett, D.} \& \bibinfo{author}{Remez, B.}
\newblock \bibinfo{title}{On electrically tunable stacking domains and ferroelectricity in moir{\'e} superlattices}.
\newblock \emph{\bibinfo{journal}{npj 2{D} Mater. Appl.}} \textbf{\bibinfo{volume}{6}}, \bibinfo{pages}{1--6} (\bibinfo{year}{2022}).
\newblock \urlprefix\url{https://doi.org/10.1038/s41699-021-00281-6}.

\bibitem{bennett2022theory}
\bibinfo{author}{Bennett, D.}
\newblock \bibinfo{title}{Theory of polar domains in moiré heterostructures}.
\newblock \emph{\bibinfo{journal}{Phys. Rev. B}} \textbf{\bibinfo{volume}{105}}, \bibinfo{pages}{235445} (\bibinfo{year}{2022}).
\newblock \urlprefix\url{https://doi.org/10.1103/PhysRevB.105.235445}.

\bibitem{bennett2023polar}
\bibinfo{author}{Bennett, D.}, \bibinfo{author}{Chaudhary, G.}, \bibinfo{author}{Slager, R.-J.}, \bibinfo{author}{Bousquet, E.} \& \bibinfo{author}{Ghosez, P.}
\newblock \bibinfo{title}{Polar meron-antimeron networks in strained and twisted bilayers}.
\newblock \emph{\bibinfo{journal}{Nat. Commun.}} \textbf{\bibinfo{volume}{14}}, \bibinfo{pages}{1629} (\bibinfo{year}{2023}).
\newblock \urlprefix\url{https://doi.org/10.1038/s41467-023-37337-8}.

\bibitem{bennett2023theory}
\bibinfo{author}{Bennett, D.}, \bibinfo{author}{Jankowski, W.~J.}, \bibinfo{author}{Chaudhary, G.}, \bibinfo{author}{Kaxiras, E.} \& \bibinfo{author}{Slager, R.-J.}
\newblock \bibinfo{title}{Theory of polarization textures in crystal supercells}.
\newblock \emph{\bibinfo{journal}{Phys. Rev. Res.}} \textbf{\bibinfo{volume}{5}}, \bibinfo{pages}{033216} (\bibinfo{year}{2023}).
\newblock \urlprefix\url{https://link.aps.org/doi/10.1103/PhysRevResearch.5.033216}.

\bibitem{Rmp1}
\bibinfo{author}{Qi, X.-L.} \& \bibinfo{author}{Zhang, S.-C.}
\newblock \bibinfo{title}{Topological insulators and superconductors}.
\newblock \emph{\bibinfo{journal}{Rev. Mod. Phys.}} \textbf{\bibinfo{volume}{83}}, \bibinfo{pages}{1057--1110} (\bibinfo{year}{2011}).
\newblock \urlprefix\url{https://link.aps.org/doi/10.1103/RevModPhys.83.1057}.

\bibitem{Rmp2}
\bibinfo{author}{Hasan, M.~Z.} \& \bibinfo{author}{Kane, C.~L.}
\newblock \bibinfo{title}{Colloquium}.
\newblock \emph{\bibinfo{journal}{Rev. Mod. Phys.}} \textbf{\bibinfo{volume}{82}}, \bibinfo{pages}{3045--3067} (\bibinfo{year}{2010}).
\newblock \urlprefix\url{https://link.aps.org/doi/10.1103/RevModPhys.82.3045}.

\bibitem{jankowski2024polarization}
\bibinfo{author}{Jankowski, W.~J.}, \bibinfo{author}{Bennett, D.}, \bibinfo{author}{Agarwal, A.}, \bibinfo{author}{Chaudhary, G.} \& \bibinfo{author}{Slager, R.-J.}
\newblock \bibinfo{title}{Polarization textures in crystal supercells with topological bands}.
\newblock \emph{\bibinfo{journal}{Phys. Rev. B}} \textbf{\bibinfo{volume}{110}}, \bibinfo{pages}{085429} (\bibinfo{year}{2024}).
\newblock \urlprefix\url{https://link.aps.org/doi/10.1103/PhysRevB.110.085429}.

\bibitem{das2019observation}
\bibinfo{author}{Das, S.} \emph{et~al.}
\newblock \bibinfo{title}{Observation of room-temperature polar skyrmions}.
\newblock \emph{\bibinfo{journal}{Nature}} \textbf{\bibinfo{volume}{568}}, \bibinfo{pages}{368--372} (\bibinfo{year}{2019}).
\newblock \urlprefix\url{https://doi.org/10.1038/s41586-019-1092-8}.

\bibitem{han2022high}
\bibinfo{author}{Han, L.} \emph{et~al.}
\newblock \bibinfo{title}{High-density switchable skyrmion-like polar nanodomains integrated on silicon}.
\newblock \emph{\bibinfo{journal}{Nature}} \textbf{\bibinfo{volume}{603}}, \bibinfo{pages}{63--67} (\bibinfo{year}{2022}).
\newblock \urlprefix\url{https://doi.org/10.1038/s41586-021-04338-w}.

\bibitem{junquera2023topologicaly}
\bibinfo{author}{Junquera, J.} \emph{et~al.}
\newblock \bibinfo{title}{Topological phases in polar oxide nanostructures}.
\newblock \emph{\bibinfo{journal}{Rev. Mod. Phys.}} \textbf{\bibinfo{volume}{95}}, \bibinfo{pages}{025001} (\bibinfo{year}{2023}).
\newblock \urlprefix\url{https://link.aps.org/doi/10.1103/RevModPhys.95.025001}.

\bibitem{sanchez20242d}
\bibinfo{author}{S{\'a}nchez-Santolino, G.} \emph{et~al.}
\newblock \bibinfo{title}{A 2{D} ferroelectric vortex pattern in twisted {B}a{T}i{O}$_3$ freestanding layers}.
\newblock \emph{\bibinfo{journal}{Nature}} \textbf{\bibinfo{volume}{626}}, \bibinfo{pages}{529--534} (\bibinfo{year}{2024}).
\newblock \urlprefix\url{https://doi.org/10.1038/s41586-023-06978-6}.

\bibitem{vu2024imaging}
\bibinfo{author}{Vu, T.-H.-Y.} \emph{et~al.}
\newblock \bibinfo{title}{Imaging topological polar structures in marginally twisted 2{D} semiconductors}.
\newblock \emph{\bibinfo{journal}{arxiv:2405.15126}}  (\bibinfo{year}{2024}).
\newblock \urlprefix\url{https://arxiv.org/abs/2405.15126}.

\bibitem{Belinicher1982}
\bibinfo{author}{Belinicher, V.}, \bibinfo{author}{Ivchenko, E.} \& \bibinfo{author}{Sturman, B.}
\newblock \bibinfo{title}{Kinetic theory of the displacement photovoltaic effect in piezoelectric}.
\newblock \emph{\bibinfo{journal}{Sov. Phys. JETP}} \textbf{\bibinfo{volume}{56}}, \bibinfo{pages}{359} (\bibinfo{year}{1982}).
\newblock \urlprefix\url{http://jetp.ras.ru/cgi-bin/dn/e_056_02_0359.pdf}.

\bibitem{ochoa2020}
\bibinfo{author}{Ochoa, H.} \& \bibinfo{author}{Asenjo-Garcia, A.}
\newblock \bibinfo{title}{Flat bands and chiral optical response of moir\'e insulators}.
\newblock \emph{\bibinfo{journal}{Phys. Rev. Lett.}} \textbf{\bibinfo{volume}{125}}, \bibinfo{pages}{037402} (\bibinfo{year}{2020}).
\newblock \urlprefix\url{https://link.aps.org/doi/10.1103/PhysRevLett.125.037402}.

\bibitem{Hesp_2021}
\bibinfo{author}{Hesp, N.~C.} \emph{et~al.}
\newblock \bibinfo{title}{Nano-imaging photoresponse in a moir{\'e} unit cell of minimally twisted bilayer graphene}.
\newblock \emph{\bibinfo{journal}{Nat. Commun.}} \textbf{\bibinfo{volume}{12}}, \bibinfo{pages}{1640} (\bibinfo{year}{2021}).
\newblock \urlprefix\url{https://www.nature.com/articles/s41467-021-21862-5}.

\bibitem{Zhang_2023}
\bibinfo{author}{Zhang, S.} \emph{et~al.}
\newblock \bibinfo{title}{Visualizing moir{\'e} ferroelectricity via plasmons and nano-photocurrent in graphene/twisted-{WS}e$_2$ structures}.
\newblock \emph{\bibinfo{journal}{Nat. Commun.}} \textbf{\bibinfo{volume}{14}}, \bibinfo{pages}{6200} (\bibinfo{year}{2023}).
\newblock \urlprefix\url{https://www.nature.com/articles/s41467-023-41773-x}.

\bibitem{Du2023}
\bibinfo{author}{Du, L.} \emph{et~al.}
\newblock \bibinfo{title}{Moir{\'e} photonics and optoelectronics}.
\newblock \emph{\bibinfo{journal}{Science}} \textbf{\bibinfo{volume}{379}}, \bibinfo{pages}{eadg0014} (\bibinfo{year}{2023}).
\newblock \urlprefix\url{https://www.science.org/doi/10.1126/science.adg0014}.

\bibitem{kuang2024opticalpropertiesplasmonsmoire}
\bibinfo{author}{Kuang, X.} \emph{et~al.}
\newblock \bibinfo{title}{Optical properties and plasmons in moir{\'e} structures}.
\newblock \emph{\bibinfo{journal}{J. Phys.: Condens. Matter}} \textbf{\bibinfo{volume}{36}}, \bibinfo{pages}{173001} (\bibinfo{year}{2024}).
\newblock \urlprefix\url{https://doi.org/10.1088/1361-648X/ad1f8c}.

\bibitem{zhang2024plasmonic}
\bibinfo{author}{Zhang, S.} \emph{et~al.}
\newblock \bibinfo{title}{Plasmonic polarization sensing of electrostatic superlattice potentials}.
\newblock \emph{\bibinfo{journal}{arXiv:2406.18028}}  (\bibinfo{year}{2024}).
\newblock \urlprefix\url{https://arxiv.org/abs/2406.18028}.

\bibitem{Sipe1993}
\bibinfo{author}{Sipe, J.~E.} \& \bibinfo{author}{Ghahramani, E.}
\newblock \bibinfo{title}{Nonlinear optical response of semiconductors in the independent-particle approximation}.
\newblock \emph{\bibinfo{journal}{Phys. Rev. B}} \textbf{\bibinfo{volume}{48}}, \bibinfo{pages}{11705--11722} (\bibinfo{year}{1993}).
\newblock \urlprefix\url{https://link.aps.org/doi/10.1103/PhysRevB.48.11705}.

\bibitem{Sipe2000}
\bibinfo{author}{Sipe, J.~E.} \& \bibinfo{author}{Shkrebtii, A.~I.}
\newblock \bibinfo{title}{Second-order optical response in semiconductors}.
\newblock \emph{\bibinfo{journal}{Phys. Rev. B}} \textbf{\bibinfo{volume}{61}}, \bibinfo{pages}{5337--5352} (\bibinfo{year}{2000}).
\newblock \urlprefix\url{https://link.aps.org/doi/10.1103/PhysRevB.61.5337}.

\bibitem{chaudhary2022}
\bibinfo{author}{Chaudhary, S.}, \bibinfo{author}{Lewandowski, C.} \& \bibinfo{author}{Refael, G.}
\newblock \bibinfo{title}{Shift-current response as a probe of quantum geometry and electron-electron interactions in twisted bilayer graphene}.
\newblock \emph{\bibinfo{journal}{Phys. Rev. Res.}} \textbf{\bibinfo{volume}{4}}, \bibinfo{pages}{013164} (\bibinfo{year}{2022}).
\newblock \urlprefix\url{https://link.aps.org/doi/10.1103/PhysRevResearch.4.013164}.

\bibitem{chen2024enhancingshiftcurrentvirtual}
\bibinfo{author}{Chen, S.}, \bibinfo{author}{Chaudhary, S.}, \bibinfo{author}{Refael, G.} \& \bibinfo{author}{Lewandowski, C.}
\newblock \bibinfo{title}{Enhancing shift current response via virtual multiband transitions}.
\newblock \emph{\bibinfo{journal}{Commun. Phys.}} \textbf{\bibinfo{volume}{7}}, \bibinfo{pages}{250} (\bibinfo{year}{2024}).
\newblock \urlprefix\url{https://www.nature.com/articles/s42005-024-01729-z}.

\bibitem{Cook2017}
\bibinfo{author}{Cook, A.~M.}, \bibinfo{author}{M.~Fregoso, B.}, \bibinfo{author}{De~Juan, F.}, \bibinfo{author}{Coh, S.} \& \bibinfo{author}{Moore, J.~E.}
\newblock \bibinfo{title}{Design principles for shift current photovoltaics}.
\newblock \emph{\bibinfo{journal}{Nat. Commun.}} \textbf{\bibinfo{volume}{8}}, \bibinfo{pages}{14176} (\bibinfo{year}{2017}).
\newblock \urlprefix\url{https://www.nature.com/articles/ncomms14176}.

\bibitem{kaplan2022twisted}
\bibinfo{author}{Kaplan, D.}, \bibinfo{author}{Holder, T.} \& \bibinfo{author}{Yan, B.}
\newblock \bibinfo{title}{Twisted photovoltaics at terahertz frequencies from momentum shift current}.
\newblock \emph{\bibinfo{journal}{Phys. Rev. Res.}} \textbf{\bibinfo{volume}{4}}, \bibinfo{pages}{013209} (\bibinfo{year}{2022}).
\newblock \urlprefix\url{https://link.aps.org/doi/10.1103/PhysRevResearch.4.013209}.

\bibitem{zhu2024anomalousshiftopticalvorticity}
\bibinfo{author}{Zhu, P.} \& \bibinfo{author}{Alexandradinata, A.}
\newblock \bibinfo{title}{Anomalous shift and optical vorticity in the steady photovoltaic current}.
\newblock \emph{\bibinfo{journal}{Phys. Rev. B}} \textbf{\bibinfo{volume}{110}}, \bibinfo{pages}{115108} (\bibinfo{year}{2024}).
\newblock \urlprefix\url{https://link.aps.org/doi/10.1103/PhysRevB.110.115108}.

\bibitem{bouhon2023quantum}
\bibinfo{author}{Bouhon, A.}, \bibinfo{author}{Timmel, A.} \& \bibinfo{author}{Slager, R.-J.}
\newblock \bibinfo{title}{Quantum geometry beyond projective single bands}.
\newblock \emph{\bibinfo{journal}{arXiv:2303.02180}}  (\bibinfo{year}{2023}).
\newblock \urlprefix\url{https://arxiv.org/abs/2303.02180}.

\bibitem{tormaessay}
\bibinfo{author}{T\"orm\"a, P.}
\newblock \bibinfo{title}{Essay: Where can quantum geometry lead us?}
\newblock \emph{\bibinfo{journal}{Phys. Rev. Lett.}} \textbf{\bibinfo{volume}{131}}, \bibinfo{pages}{240001} (\bibinfo{year}{2023}).
\newblock \urlprefix\url{https://link.aps.org/doi/10.1103/PhysRevLett.131.240001}.

\bibitem{Ahn2021}
\bibinfo{author}{Ahn, J.}, \bibinfo{author}{Guo, G.-Y.}, \bibinfo{author}{Nagaosa, N.} \& \bibinfo{author}{Vishwanath, A.}
\newblock \bibinfo{title}{Riemannian geometry of resonant optical responses}.
\newblock \emph{\bibinfo{journal}{Nat. Phys.}} \textbf{\bibinfo{volume}{18}}, \bibinfo{pages}{290--295} (\bibinfo{year}{2021}).
\newblock \urlprefix\url{https://doi.org/10.1038%2Fs41567-021-01465-z}.

\bibitem{provost1980riemannian}
\bibinfo{author}{Provost, J.} \& \bibinfo{author}{Vallee, G.}
\newblock \bibinfo{title}{Riemannian structure on manifolds of quantum states}.
\newblock \emph{\bibinfo{journal}{Commun. Math. Phys.}} \textbf{\bibinfo{volume}{76}}, \bibinfo{pages}{289--301} (\bibinfo{year}{1980}).
\newblock \urlprefix\url{https://doi.org/10.1007/BF02193559}.

\bibitem{vanderbilt2018berry}
\bibinfo{author}{Vanderbilt, D.}
\newblock \emph{\bibinfo{title}{Berry phases in electronic structure theory: electric polarization, orbital magnetization and topological insulators}} (\bibinfo{publisher}{Cambridge University Press}, \bibinfo{year}{2018}).

\bibitem{topp2021}
\bibinfo{author}{Topp, G.~E.}, \bibinfo{author}{Eckhardt, C.~J.}, \bibinfo{author}{Kennes, D.~M.}, \bibinfo{author}{Sentef, M.~A.} \& \bibinfo{author}{T\"orm\"a, P.}
\newblock \bibinfo{title}{Light-matter coupling and quantum geometry in moir\'e materials}.
\newblock \emph{\bibinfo{journal}{Phys. Rev. B}} \textbf{\bibinfo{volume}{104}}, \bibinfo{pages}{064306} (\bibinfo{year}{2021}).
\newblock \urlprefix\url{https://link.aps.org/doi/10.1103/PhysRevB.104.064306}.

\bibitem{king1993theory}
\bibinfo{author}{King-Smith, R.} \& \bibinfo{author}{Vanderbilt, D.}
\newblock \bibinfo{title}{Theory of polarization of crystalline solids}.
\newblock \emph{\bibinfo{journal}{Phys. Rev. B}} \textbf{\bibinfo{volume}{47}}, \bibinfo{pages}{1651} (\bibinfo{year}{1993}).
\newblock \urlprefix\url{https://doi.org/10.1103/PhysRevB.47.1651}.

\bibitem{vanderbilt1993electric}
\bibinfo{author}{Vanderbilt, D.} \& \bibinfo{author}{King-Smith, R.}
\newblock \bibinfo{title}{Electric polarization as a bulk quantity and its relation to surface charge}.
\newblock \emph{\bibinfo{journal}{Phys. Rev. B}} \textbf{\bibinfo{volume}{48}}, \bibinfo{pages}{4442} (\bibinfo{year}{1993}).
\newblock \urlprefix\url{https://doi.org/10.1103/PhysRevB.48.4442}.

\bibitem{resta1994macroscopic}
\bibinfo{author}{Resta, R.}
\newblock \bibinfo{title}{Macroscopic polarization in crystalline dielectrics: the geometric phase approach}.
\newblock \emph{\bibinfo{journal}{Rev. Mod. Phys.}} \textbf{\bibinfo{volume}{66}}, \bibinfo{pages}{899--915} (\bibinfo{year}{1994}).
\newblock \urlprefix\url{https://link.aps.org/doi/10.1103/RevModPhys.66.899}.

\bibitem{Ma2023}
\bibinfo{author}{Ma, Q.}, \bibinfo{author}{Krishna~Kumar, R.}, \bibinfo{author}{Xu, S.-Y.}, \bibinfo{author}{Koppens, F. H.~L.} \& \bibinfo{author}{Song, J. C.~W.}
\newblock \bibinfo{title}{Photocurrent as a multiphysics diagnostic of quantum materials}.
\newblock \emph{\bibinfo{journal}{Nat. Rev. Phys.}} \textbf{\bibinfo{volume}{5}}, \bibinfo{pages}{170--184} (\bibinfo{year}{2023}).
\newblock \urlprefix\url{https://doi.org/10.1038/s42254-022-00551-2}.

\bibitem{Chen2023}
\bibinfo{author}{Hu, C.}, \bibinfo{author}{Naik, M.~H.}, \bibinfo{author}{Chan, Y.-H.}, \bibinfo{author}{Ruan, J.} \& \bibinfo{author}{Louie, S.~G.}
\newblock \bibinfo{title}{Light-induced shift current vortex crystals in moiré heterobilayers}.
\newblock \emph{\bibinfo{journal}{PNAS}} \textbf{\bibinfo{volume}{120}}, \bibinfo{pages}{e2314775120} (\bibinfo{year}{2023}).
\newblock \urlprefix\url{https://www.pnas.org/doi/abs/10.1073/pnas.2314775120}.

\bibitem{Fregoso2017}
\bibinfo{author}{Fregoso, B.~M.}, \bibinfo{author}{Morimoto, T.} \& \bibinfo{author}{Moore, J.~E.}
\newblock \bibinfo{title}{Quantitative relationship between polarization differences and the zone-averaged shift photocurrent}.
\newblock \emph{\bibinfo{journal}{Phys. Rev. B}} \textbf{\bibinfo{volume}{96}}, \bibinfo{pages}{075421} (\bibinfo{year}{2017}).
\newblock \urlprefix\url{https://link.aps.org/doi/10.1103/PhysRevB.96.075421}.

\bibitem{resta2024geo}
\bibinfo{author}{Resta, R.}
\newblock \bibinfo{title}{Geometrical theory of the shift current in presence of disorder and interaction}.
\newblock \emph{\bibinfo{journal}{Phys. Rev. Lett.}} \textbf{\bibinfo{volume}{133}}, \bibinfo{pages}{206903} (\bibinfo{year}{2024}).
\newblock \urlprefix\url{https://link.aps.org/doi/10.1103/PhysRevLett.133.206903}.

\bibitem{Carr2020review}
\bibinfo{author}{Carr, S.}, \bibinfo{author}{Fang, S.} \& \bibinfo{author}{Kaxiras, E.}
\newblock \bibinfo{title}{Electronic-structure methods for twisted moiré layers}.
\newblock \emph{\bibinfo{journal}{Nature Reviews Materials}} \textbf{\bibinfo{volume}{5}}, \bibinfo{pages}{748–763} (\bibinfo{year}{2020}).
\newblock \urlprefix\url{http://dx.doi.org/10.1038/s41578-020-0214-0}.

\bibitem{carr2018relaxation}
\bibinfo{author}{Carr, S.} \emph{et~al.}
\newblock \bibinfo{title}{Relaxation and domain formation in incommensurate two-dimensional heterostructures}.
\newblock \emph{\bibinfo{journal}{Phys. Rev. B}} \textbf{\bibinfo{volume}{98}}, \bibinfo{pages}{224102} (\bibinfo{year}{2018}).
\newblock \urlprefix\url{https://doi.org/10.1103/PhysRevB.98.224102}.

\bibitem{Ahn2020}
\bibinfo{author}{Ahn, J.}, \bibinfo{author}{Guo, G.-Y.} \& \bibinfo{author}{Nagaosa, N.}
\newblock \bibinfo{title}{Low-frequency divergence and quantum geometry of the bulk photovoltaic effect in topological semimetals}.
\newblock \emph{\bibinfo{journal}{Phys. Rev. X}} \textbf{\bibinfo{volume}{10}}, \bibinfo{pages}{041041} (\bibinfo{year}{2020}).
\newblock \urlprefix\url{https://link.aps.org/doi/10.1103/PhysRevX.10.041041}.

\bibitem{marzari2012maximally}
\bibinfo{author}{Marzari, N.}, \bibinfo{author}{Mostofi, A.~A.}, \bibinfo{author}{Yates, J.~R.}, \bibinfo{author}{Souza, I.} \& \bibinfo{author}{Vanderbilt, D.}
\newblock \bibinfo{title}{Maximally localized {W}annier functions: {T}heory and applications}.
\newblock \emph{\bibinfo{journal}{Rev. Mod. Phys.}} \textbf{\bibinfo{volume}{84}}, \bibinfo{pages}{1419--1475} (\bibinfo{year}{2012}).
\newblock \urlprefix\url{https://link.aps.org/doi/10.1103/RevModPhys.84.1419}.

\bibitem{pizzi2020wannier90}
\bibinfo{author}{Pizzi, G.} \emph{et~al.}
\newblock \bibinfo{title}{Wannier90 as a community code: new features and applications}.
\newblock \emph{\bibinfo{journal}{J. Phys.: Condens. Matter}} \textbf{\bibinfo{volume}{32}}, \bibinfo{pages}{165902} (\bibinfo{year}{2020}).
\newblock \urlprefix\url{https://doi.org/10.1088/1361-648X/ab51ff}.

\bibitem{Ibanez2018}
\bibinfo{author}{Iba\~nez Azpiroz, J.}, \bibinfo{author}{Tsirkin, S.~S.} \& \bibinfo{author}{Souza, I.}
\newblock \bibinfo{title}{Ab initio calculation of the shift photocurrent by {W}annier interpolation}.
\newblock \emph{\bibinfo{journal}{Phys. Rev. B}} \textbf{\bibinfo{volume}{97}}, \bibinfo{pages}{245143} (\bibinfo{year}{2018}).
\newblock \urlprefix\url{https://link.aps.org/doi/10.1103/PhysRevB.97.245143}.

\bibitem{ji2023general}
\bibinfo{author}{Ji, J.}, \bibinfo{author}{Yu, G.}, \bibinfo{author}{Xu, C.} \& \bibinfo{author}{Xiang, H.~J.}
\newblock \bibinfo{title}{General theory for bilayer stacking ferroelectricity}.
\newblock \emph{\bibinfo{journal}{Phys. Rev. Lett.}} \textbf{\bibinfo{volume}{130}}, \bibinfo{pages}{146801} (\bibinfo{year}{2023}).
\newblock \urlprefix\url{https://link.aps.org/doi/10.1103/PhysRevLett.130.146801}.

\bibitem{lang1978scanning}
\bibinfo{author}{Lang, D.~V.} \& \bibinfo{author}{Henry, C.~H.}
\newblock \bibinfo{title}{Scanning photocurrent microscopy: A new technique to study inhomogeneously distributed recombination centers in semiconductors}.
\newblock \emph{\bibinfo{journal}{Solid-State Electron.}} \textbf{\bibinfo{volume}{21}}, \bibinfo{pages}{1519--1524} (\bibinfo{year}{1978}).
\newblock \urlprefix\url{https://www.sciencedirect.com/science/article/pii/0038110178902356}.

\bibitem{rauhut2012antenna}
\bibinfo{author}{Rauhut, N.} \emph{et~al.}
\newblock \bibinfo{title}{Antenna-enhanced photocurrent microscopy on single-walled carbon nanotubes at 30 nm resolution}.
\newblock \emph{\bibinfo{journal}{ACS Nano}} \textbf{\bibinfo{volume}{6}}, \bibinfo{pages}{6416--6421} (\bibinfo{year}{2012}).
\newblock \urlprefix\url{https://pubs.acs.org/doi/10.1021/nn301979c}.

\bibitem{Xiang2024}
\bibinfo{author}{Xiang, L.}, \bibinfo{author}{Jin, H.} \& \bibinfo{author}{Wang, J.}
\newblock \bibinfo{title}{Quantifying the photocurrent fluctuation in quantum materials by shot noise}.
\newblock \emph{\bibinfo{journal}{Nature Communications}} \textbf{\bibinfo{volume}{15}}, \bibinfo{pages}{2012} (\bibinfo{year}{2024}).
\newblock \urlprefix\url{https://doi.org/10.1038/s41467-024-46264-1}.

\bibitem{li2024imaging}
\bibinfo{author}{Li, H.} \emph{et~al.}
\newblock \bibinfo{title}{Imaging moir{\'e} excited states with photocurrent tunnelling microscopy}.
\newblock \emph{\bibinfo{journal}{Nat. Mater.}} \textbf{\bibinfo{volume}{23}}, \bibinfo{pages}{633--638} (\bibinfo{year}{2024}).
\newblock \urlprefix\url{https://doi.org/10.1038/s41563-023-01753-4}.

\bibitem{Sotome2019}
\bibinfo{author}{Sotome, M.} \emph{et~al.}
\newblock \bibinfo{title}{Spectral dynamics of shift current in ferroelectric semiconductor sbsi}.
\newblock \emph{\bibinfo{journal}{PNAS}} \textbf{\bibinfo{volume}{116}}, \bibinfo{pages}{1929--1933} (\bibinfo{year}{2019}).
\newblock \urlprefix\url{https://www.pnas.org/doi/abs/10.1073/pnas.1802427116}.

\bibitem{Yu2023}
\bibinfo{author}{Yu, H.}, \bibinfo{author}{Zhou, Z.} \& \bibinfo{author}{Yao, W.}
\newblock \bibinfo{title}{Distinct moir{\'e} textures of in-plane electric polarizations for distinguishing moir{\'e} origins in homobilayers}.
\newblock \emph{\bibinfo{journal}{Sci. China Phys. Mech. Astron.}} \textbf{\bibinfo{volume}{66}}, \bibinfo{pages}{107711} (\bibinfo{year}{2023}).
\newblock \urlprefix\url{https://doi.org/10.1007/s11433-023-2163-3}.

\bibitem{gonze2016}
\bibinfo{author}{Gonze, X.} \& \bibinfo{author}{et~al.}
\newblock \bibinfo{title}{Recent developments in the {ABINIT} software package}.
\newblock \emph{\bibinfo{journal}{Comput. Phys. Commun.}} \textbf{\bibinfo{volume}{205}}, \bibinfo{pages}{106 -- 131} (\bibinfo{year}{2016}).
\newblock \urlprefix\url{https://www.sciencedirect.com/science/article/pii/S0010465516300923}.

\bibitem{gonze2020}
\bibinfo{author}{Gonze, X.} \& \bibinfo{author}{et~al.}
\newblock \bibinfo{title}{The {ABINIT} project: Impact, environment and recent developments}.
\newblock \emph{\bibinfo{journal}{Comput. Phys. Commun.}} \textbf{\bibinfo{volume}{248}}, \bibinfo{pages}{107042} (\bibinfo{year}{2020}).
\newblock \urlprefix\url{https://www.sciencedirect.com/science/article/pii/S0010465519303741}.

\bibitem{norm_conserving}
\bibinfo{author}{Hamann, D.}
\newblock \bibinfo{title}{Optimized norm-conserving {V}anderbilt pseudopotentials}.
\newblock \emph{\bibinfo{journal}{Phys. Rev. B}} \textbf{\bibinfo{volume}{88}}, \bibinfo{pages}{085117} (\bibinfo{year}{2013}).
\newblock \urlprefix\url{https://doi.org/10.1103/PhysRevB.88.085117}.

\bibitem{psml}
\bibinfo{author}{Garc{\'\i}a, A.}, \bibinfo{author}{Verstraete, M.~J.}, \bibinfo{author}{Pouillon, Y.} \& \bibinfo{author}{Junquera, J.}
\newblock \bibinfo{title}{The {PSML} format and library for norm-conserving pseudopotential data curation and interoperability}.
\newblock \emph{\bibinfo{journal}{Comput. Phys. Commun.}} \textbf{\bibinfo{volume}{227}}, \bibinfo{pages}{51} (\bibinfo{year}{2018}).
\newblock \urlprefix\url{https://doi.org/10.1016/j.cpc.2018.02.011}.

\bibitem{pseudodojo}
\bibinfo{author}{Van~Setten, M.} \emph{et~al.}
\newblock \bibinfo{title}{The pseudodojo: Training and grading a 85 element optimized norm-conserving pseudopotential table}.
\newblock \emph{\bibinfo{journal}{Comput. Phys. Commun.}} \textbf{\bibinfo{volume}{226}}, \bibinfo{pages}{39} (\bibinfo{year}{2018}).
\newblock \urlprefix\url{https://doi.org/10.1016/j.cpc.2018.01.012}.

\bibitem{mp}
\bibinfo{author}{Monkhorst, H.~J.} \& \bibinfo{author}{Pack, J.~D.}
\newblock \bibinfo{title}{Special points for {B}rillouin-zone integrations}.
\newblock \emph{\bibinfo{journal}{Phys. Rev. B}} \textbf{\bibinfo{volume}{13}}, \bibinfo{pages}{5188} (\bibinfo{year}{1976}).
\newblock \urlprefix\url{https://doi.org/10.1103/PhysRevB.13.5188}.

\bibitem{zhang1998comment}
\bibinfo{author}{Zhang, Y.} \& \bibinfo{author}{Yang, W.}
\newblock \bibinfo{title}{Comment on “{G}eneralized gradient approximation made simple”}.
\newblock \emph{\bibinfo{journal}{Phys. Rev. Lett.}} \textbf{\bibinfo{volume}{80}}, \bibinfo{pages}{890} (\bibinfo{year}{1998}).
\newblock \urlprefix\url{https://journals.aps.org/prl/abstract/10.1103/PhysRevLett.80.890}.

\bibitem{becke2006simple}
\bibinfo{author}{Becke, A.~D.} \& \bibinfo{author}{Johnson, E.~R.}
\newblock \bibinfo{title}{A simple effective potential for exchange}.
\newblock \emph{\bibinfo{journal}{J. Chem. Phys.}} \textbf{\bibinfo{volume}{124}}, \bibinfo{pages}{221101} (\bibinfo{year}{2006}).
\newblock \urlprefix\url{https://pubs.aip.org/aip/jcp/article/124/22/221101/920551/A-simple-effective-potential-for-exchange}.

\end{thebibliography}

\clearpage



\section*{Supplementary material}

# SUPPLEMENTARY MATERIAL
## Shift photocurrent vortices from topological polarization textures

Aneesh Agarwal,[1] Wojciech J. Jankowski,[1] Daniel Bennett,[2] and Robert-Jan Slager[1]

[1]*Theory of Condensed Matter Group, Cavendish Laboratory, University of Cambridge, J. J. Thomson Avenue, Cambridge CB3 0HE, UK*
[2]*John A. Paulson School of Engineering and Applied Sciences, Harvard University, Cambridge, Massachusetts 02138, USA*

## I. MORE DETAILS ON QUANTUM GEOMETRY

We provide further details on the quantum geometry relevant to this work [S1] . The quantum-geometric tensor (QGT) $Q_{mn}^{ab}$ was defined in terms of non-Abelian Berry connection $A_{nm}^a(\mathbf{k}) = i \langle u_{n,\mathbf{k}} | \partial_{k_a} u_{m,\mathbf{k}} \rangle$, with spatial indices $a = x, y, \ldots$, as:

$$Q_{mn}^{ab} \equiv A_{nm}^a A_{mn}^b = g_{mn}^{ab} - \frac{i}{2} F_{mn}^{ab},$$ (S1)

which here we furthermore decompose into the quantum metric $g_{mn}^{ab}$ (real part) and Berry curvature form contributions $F_{mn}^{ab}$ [S2] (imaginary part); the Berry curvature in band $n$ then reads: $\Omega_{ab}^n = \sum_m^{\text{unocc}} F_{mn}^{ab}$. Notably, in the context of the main text, the shift photoconductivity components $\sigma^{c,xx}$ and $\sigma^{c,yy}$ only couple electric fields to the quantum metric $g_{mn}^{xx}$ and $g_{mn}^{yy}$ through the polarization-related shift vector part, whereas the components $\sigma^{x,xy}\sigma^{x,yx}, \sigma^{y,xy}, \sigma^{y,yx}$ involve coupling to both the quantum metric, $g_{mn}^{xy} = g_{mn}^{yx}$, and symplectic elements, $F_{mn}^{xy} = -F_{mn}^{yx}$.

Following Ref. [S3], we can further formalize the formulation of the QGT-based objects that are relevant for studying the considered optical responses. Accordingly, one can define a transition dipole moment, $\mathbf{r}_{mn}(\mathbf{k}) = \langle \psi_{m\mathbf{k}} | \hat{\mathbf{r}} | \psi_{n\mathbf{k}} \rangle$, and identify it as a tangent vector on the manifold of quantum states with components [S3],

$$r_{mn}^a(\mathbf{k}) = i\delta_{mn}\partial_a + A_{mn}^a(\mathbf{k}).$$ (S2)

Here, $\partial_a \equiv \frac{\partial}{\partial_{k_a}}$ is a tangent vector component induced by the local parameter-space coordinates $\{k_a\}$. Furthermore, local interband transition vectors can be defined [S3],

$$\hat{e}_{mn}^a(\mathbf{k}) = \mathbf{r}_{mn}^a(\mathbf{k}) |u_{m,\mathbf{k}}\rangle \langle u_{n,\mathbf{k}}| \,,$$ (S3)

which rigorously define the QGT as a Hermitian metric induced by an inner product of the tangent vectors,

$$Q_{mn}^{ab} \equiv \langle \hat{e}_{mn}^a, \hat{e}_{mn}^b \rangle = r_{nm}^a r_{mn}^b.$$ (S4)

Here, a Hilbert-Schmidt inner product of matrices is used, which is defined as [S3],

$$\langle A, B \rangle = \text{Tr}[A^\dagger B] = \sum_{a,b} A_{ab}^* B_{ab}.$$ (S5)

As relevant to the shift currents, we further define a Hermitian connection, consistently with Ref. [S3],

$$C_{mn}^{acb} = \langle \hat{e}_{mn}^a, \mathcal{D}_c \hat{e}_{mn}^b \rangle = r_{nm}^a r_{mn}^{b;c},$$ (S6)

where a covariant derivative can be defined with a parallel-transport equation $\mathcal{D}_c \hat{e}_{mn}^a = \sum_b (C_{mn}^{ca})_b \hat{e}_{mn}^b$ and $C_{mn}^{acb} \equiv \sum_d Q_{mn}^{ad}(C_{mn}^{cb})_d$. Here, $r_{nm}^{b;c}$ is a generalized (covariant) derivative of the transition dipole,

$$r_{mn}^{b;c}(\mathbf{k}) \equiv \mathcal{D}_c r_{mn}^b(\mathbf{k}) = \partial_c r_{mn}^b(\mathbf{k}) - i[A_{mm}^c(\mathbf{k}) - A_{nn}^c(\mathbf{k})]r_{mn}^b(\mathbf{k}),$$ (S7)

as utilized in the Methods.

## II. OPTICAL RESPONSES

The optical responses relevant to this work can be recast in terms of their quantum-geometric interpretations, following Refs. [S2, S3] which can be derived from perturbation theory in electric dipole Hamiltonian, for more details see Refs. [S4, S5].

The geometric quantities introduced in the previous Section are physically related to the **first-order optical conductivity**

$$\sigma^{b,a} = \frac{\pi\omega e^2}{\hbar} \sum_{m,n} \int \frac{\mathrm{d}^d \mathbf{k}}{(2\pi)^d} \delta(\omega - \omega_{mn}) f_{nm} Q_{mn}^{ba},$$ (S8)

where $f_{nm} = f_{n,\mathbf{k}} - f_{m,\mathbf{k}}$ is the difference between the Fermi-Dirac probability distributions of (occupied) bands $n$ and (unoccupied) bands $m$. $\hbar\omega_{mn}$ is the energy gap as a function of $\mathbf{k}$ between the bands $m$ and $n$ involved in the photoexcitation.

Furthermore, the quantum-geometric quantities can be used to calculate second-order photoconductivities in materials. Such photoconductivities can yield DC photovoltaic responses, which for a second-order response decompose into injection and shift terms,

$$\sigma^{c,ab} = \sigma_{\text{inj}}^{c,ab} + \sigma_{\text{shift}}^{c,ab}. \tag{S9}$$

The injection currents arise due to the different band dispersions of the valence and conduction, which cause the group velocities of the photoexcited electrons to change. The associated **injection photoconductivity** can be written in terms of the quantum geometric tensor $Q_{mn}^{ba}$ as

$$\sigma_{\text{inj}}^{c,ab} = -\frac{\pi e^3 \tau}{\hbar^2} \sum_{m,n} \int \frac{\mathrm{d}^2\mathbf{k}}{(2\pi)^2} \delta(\omega - \omega_{mn}) f_{nm} Q_{mn}^{ab} \partial_c \omega_{mn}. \tag{S10}$$

Here, $\tau$ is a relaxation time for the photoexcited particle to decay.

Centrally to this work, the **shift photoconductivity** – which is the other part of the second-order photoconductivity that yields another photovoltaic DC response to light (due to the positional shifts of electrons) beyond the injection currents – can be related to the Hermitian connection $C_{acb}^{mn}$ as [S3]

$$\sigma_{\text{shift}}^{c,ab} = -\frac{\pi e^3}{2\hbar^2} \sum_{m,n} \int \frac{\mathrm{d}^2\mathbf{k}}{(2\pi)^2} \delta(\omega - \omega_{mn}) f_{nm} i\left(C_{mn}^{acb} - (C_{mn}^{bca})^*\right). \tag{S11}$$

Crucially, the shift conductivities can also be expressed with the **shift vectors**,

$$R_{mn}^{c,ab} = A_{mm}^c - A_{nn}^c + i\partial_c \log\left(r_{nm}^a/r_{mn}^b\right), \tag{S12}$$

corresponding to positional shifts of electronic wavepackets under optical transitions [S4, S5], as can be captured by the relation of the Abelian Berry connection ($A_{nn}^c$) to Wannier charge centers [S6]. Furthermore, on direct substitution, we have,

$$\sigma_{\text{shift}}^{c,ab} = -\frac{\pi e^3}{\hbar^2} \sum_{m,n} \int \frac{\mathrm{d}^2\mathbf{k}}{(2\pi)^d} \delta(\omega - \omega_{mn}) f_{nm} Q_{mn}^{ba} R_{mn}^{c,ab} , \tag{S13}$$

which explicitly demonstrates that the non-vanishing shift photoconductivities do not require any band dispersion, unlike the injection photoconductivities.

The introduced photoconductivities capture the light-induced DC photovoltaic responses [S2, S3]:

$$j_{\text{shift/inj}}^c(0) = 2 \sum_{a,b} \sigma_{\text{shift/inj}}^{c,ab}(\omega) \mathcal{E}^a(\omega) \mathcal{E}^b(-\omega), \tag{S14}$$

for the second order DC currents ('0' denotes the vanishing frequency of the shift/injection photocurrent) induced by the electric fields $\mathcal{E}^a(\omega)$. $a, b$ run over spatial direction indices $(x, y, \ldots)$. In the main text, we drop the implicit indices capturing the vanishing frequency of the DC current, as well as its shift nature, given the context and focus of this work on these aspects.

For the first-order AC responses,

$$j^b(\omega) = \sum_a \sigma^{b,a}(\omega) \mathcal{E}^a(\omega), \tag{S15}$$

which were further calculated in the subsequent Sections to contrast with the second-order results. Importantly, the first-order responses result only in AC currents that can be associated with an oscillatory dynamical polarization of the system, whereas the second-order DC shift and injection photoconductivities result in non-oscillatory bulk currents measurable as a steady response.

### III. CONFIGURATION SPACE APPROXIMATION

We briefly elaborate on the configuration space approximation utilized in the tight-binding models and first-principles calculations performed in this work. While the configuration space approximation was used to deduce the local polarization $\mathbf{P}(\mathbf{x})$ in moiré supercells [S7–S9], its application to compute the local shift photoconductivities is a central component of this work.

We first consider a moiré bilayer with a small twist angle. For a moiré bilayer with relative twist angle $\theta$ between the layers, the mapping to configuration space is given by [S10]

$$\mathbf{x}(\mathbf{r}) = (\mathbb{I} - R_\theta^{-1})\mathbf{r} \,, \tag{S16}$$

where $R_\theta$ is a rotation matrix and $\mathbf{r}$ is the real space position modulo any lattice vector. While Eq. (S16) is exact, for small $\theta$, the local environment changes around any unit cells are small, hence the local properties in each can be faithfully approximated by a commensurate bilayer with a relative translation between the layers. The translation can be expressed as

$$\mathbf{x} \approx \theta \begin{bmatrix} 0 & -1 \\ 1 & 0 \end{bmatrix} \mathbf{r}. \tag{S17}$$

allowing the local properties in small-angle twisted bilayers to be parameterized efficiently using a single commensurate cell of a bilayer, and sliding one layer over the other, when performing first-principles calculations. What we define as the cell at $\mathbf{r}_j$ for moiré systems, is a unit cell of one layer *and* the atoms of the other layer contained (upon a projection) in that unit cell. A general projective description of this kind is natural for the configuration space picture, providing an arena to define a local quantity, such as local polarization [S8].

## IV. RECONSTRUCTING LOCAL ELECTRIC POLARIZATION FROM THE SHIFT CURRENTS

We further detail the reconstruction scheme for the in-plane electric polarization, based on the shift photocurrents present. The shift photoconductivity component in quadratic response to electric field in the $x$-direction reads,

$$\sigma^{c,xx}(\omega) = -\frac{\pi e^3}{\hbar^2} \sum_{m,n} \int_{\mathrm{BZ}} \frac{\mathrm{d}^2 \mathbf{k}}{(2\pi)^2} \delta(\omega - \omega_{mn}) f_{nm} \, R_{mn}^c |A_{nm}^x|^2 = -\frac{\pi e^3}{\hbar^2} \sum_{m,n} \int_{\mathrm{BZ}} \frac{\mathrm{d}^2 \mathbf{k}}{(2\pi)^2} \delta(\omega - \omega_{mn}) f_{nm} \, R_{mn}^c Q_{mn}^{xx}. \tag{S18}$$

Note that here, an appropriate gauge has been chosen such that the spatial indices on the shift vector corresponding to the stimulating electric field polarizations can be dropped (as detailed in Methods and in the further Sections). Hence, on combining with an analogous component for the response in the $y$-direction,

$$\sigma^{c,xx}(\omega) + \sigma^{c,yy}(\omega) = -\frac{\pi e^3}{\hbar^2} \sum_{m,n} \int_{\mathrm{BZ}} \frac{\mathrm{d}^2 \mathbf{k}}{(2\pi)^2} \delta(\omega - \omega_{mn}) f_{nm} \, R_{mn}^c (Q_{mn}^{xx} + Q_{mn}^{yy}). \tag{S19}$$

Assuming that $Q_{mn}^{xx} + Q_{mn}^{yy}$ contributes comparably/uniformly in the momentum-space region where the polarization is dominant within the BZ (the edge of BZ in the moiré bilayers, see the main text), we can approximate

$$\sigma^{c,xx} + \sigma^{c,yy} \approx -\frac{\pi e^3}{\hbar^2} \sum_{m,n} \langle Q_{mn}^{xx} + Q_{mn}^{yy} \rangle \int_{\mathrm{BZ}} \frac{\mathrm{d}^2 \mathbf{k}}{(2\pi)^2} \delta(\omega - \omega_{mn}) f_{nm} \, R_{mn}^c, \tag{S20}$$

which follows the idea of Ref. [S11]. Additionally, within the effective tight-binding models for the moiré bilayers (see the following Sections), and owing to the triviality of the Bloch bundle over BZ, we have:

$$R_{mn}^c \approx A_{mm}^c - A_{nn}^c \approx -2A_{nn}^c. \tag{S21}$$

Therefore, on direct substitution, we further obtain,

$$\sigma^{c,xx} + \sigma^{c,yy} \approx \frac{2\pi e^3}{\hbar^2} \sum_{m,n} \langle Q_{mn}^{xx} + Q_{mn}^{yy} \rangle f_{nm} \int \frac{\mathrm{d}^2 \mathbf{k}}{(2\pi)^2} A_{nn}^c \approx -\Delta P^c \left( \frac{\pi e^2}{\hbar^2} \sum_{m,n} \langle Q_{mn}^{xx} + Q_{mn}^{yy} \rangle \right) \propto -\Delta P^c, \tag{S22}$$

with $c = x, y$, and the polarization components $\Delta P^c$ defined consistently with the modern theory of polarization [S12], as in the main text. In particular, here $P^c \equiv \Delta P^c$ is defined as a change of polarization with respect to a non-polar, or high-symmetry, state [S8]. Note that in the second equality above, it has been assumed that the two valence bands contribute roughly equally to the polarization, consistently with the model. This concludes the derivation of the correspondence between the local shift currents and electric in-plane polarizations.

## V. SYMMETRY ANALYSIS OF SHIFT PHOTOCONDUCTIVITIES

We present a symmetry analysis of the stacking-dependent shift photoconductivities. Given the correspondence between electric polarization and shift photocurrents derived above, the analysis parallels the symmetry analysis of the local in-plane electric polarization in Ref. [S7].

We choose the vector $\hat{n}$ to point along the diagonal of a supercell. Along the $\hat{n}$ direction, the order of configuration space stackings follows as: AA, AB, DW, BA, AA. According to Ref. [S7], the local polarization satisfies,

$$\mathbf{P}(\mathbf{x}) \cdot \hat{n} = \mathbf{P}(-\mathbf{x}) \cdot \hat{n}. \tag{S23}$$

under the present mirror symmetry $m : \mathbf{x} \to -\mathbf{x}$. Analogously, the shift photoconductivity vectors, as defined in the main text, satisfy,

$$\sigma(\mathbf{x}) \cdot \hat{n} = \sigma(-\mathbf{x}) \cdot \hat{n}. \tag{S24}$$

Moreover, under $C_3$ symmetry present in t-hBN, the following conditions are satisfied

$$\sigma^{x,xx}(\mathbf{x}) = -\sigma^{x,yy}(\mathbf{x}) = -\sigma^{y,xx}(\mathbf{x}) = -\sigma^{y,xy}(\mathbf{x}), \tag{S25}$$

which in the context of transition metal dichalcogenides (TMDs) was also identified in Ref. [S13].

## VI. EFFECTIVE TIGHT-BINDING MODELS

For completeness, we further detail the form of the tight-binding hopping elements used in the tight-binding Hamiltonian detailed in Methods. The interlayer coupling constants used in the model, as also defined in Methods, read [S9, S14],

$$
\begin{aligned}
t_{vc,\mathbf{k}}^{tb}(\mathbf{x}) &= \left( \alpha_{\mathbf{k}} \beta_{\mathbf{k}}^* [t_{B_b B_t}(\mathbf{x}) - t_{N_b N_t}(\mathbf{x})] + \beta_{\mathbf{k}}^{*2} \sum_{i=1}^{3} \left( t_{N_b B_t,i}(\mathbf{x}) e^{-i\mathbf{k} \cdot \Delta \mathbf{R}_i} \right) - \alpha_{\mathbf{k}}^2 \sum_{i=1}^{3} \left( t_{B_b N_t,i}(\mathbf{x}) e^{i\mathbf{k} \cdot \Delta \mathbf{R}_i} \right) \right) e^{i\mathbf{k} \cdot \mathbf{x}}, \\
t_{vc,\mathbf{k}}^{bt}(\mathbf{x}) &= \left( \alpha_{\mathbf{k}} \beta_{\mathbf{k}}^* [t_{B_b B_t}^*(\mathbf{x}) - t_{N_b N_t}^*(\mathbf{x})] + \beta_{\mathbf{k}}^{*2} \sum_{i=1}^{3} \left( t_{B_b N_t,i}^*(\mathbf{x}) e^{-i\mathbf{k} \cdot \Delta \mathbf{R}_i} \right) - \alpha_{\mathbf{k}}^2 \sum_{i=1}^{3} \left( t_{N_b B_t,i}^*(\mathbf{x}) e^{i\mathbf{k} \cdot \Delta \mathbf{R}_i} \right) \right) e^{-i\mathbf{k} \cdot \mathbf{x}}.
\end{aligned} \tag{S26}
$$

where $\Delta \mathbf{R}_i$ are displacement vectors between the atoms in the subscripts. In the above $\alpha_{\mathbf{k}}$ and $\beta_{\mathbf{k}}$ are the coefficients of the Bloch orbitals in the standard graphene/hBN Hamiltonian on a monolayer honeycomb lattice in the presence of a diagonal mass term $m\sigma_z = \text{diag}(m, -m)$ [S15]:

$$|u_{v,\mathbf{k}}\rangle = \alpha_{\mathbf{k}} \sum_i e^{i\mathbf{k} \cdot [(\mathbf{R})_i + \mathbf{r}_B]} |B_i\rangle + \beta_{\mathbf{k}} \sum_i e^{i\mathbf{k} \cdot [(\mathbf{R})_i + \mathbf{r}_N]} |N_i\rangle, \tag{S27}$$

$$|u_{c,\mathbf{k}}\rangle = (\beta_{\mathbf{k}})^* \sum_i e^{i\mathbf{k} \cdot [(\mathbf{R})_i + \mathbf{r}_B]} |B_i\rangle - \alpha_{\mathbf{k}} \sum_i e^{i\mathbf{k} \cdot [(\mathbf{R})_i + \mathbf{r}_N]} |N_i\rangle. \tag{S28}$$

Here, $(\mathbf{R})_i$ is the lattice vector of the $i$-th unit cell, $\mathbf{r}_{B/N}$ are the relative positions of boron ($B$) and nitrogen ($N$) atoms with respect to the unit cell centers, and $|B_i\rangle$, $|N_i\rangle$ denote the boron/nitrogen orbitals in the unit cell $i$, correspondingly.

The stacking-dependent hoppings can be further implemented to obtain effective eigenstates and band structures, as detailed in Methods. For completeness, we present an effective band structure obtained for the AA, AB, and DW stackings within the presented tight-binding model, see Fig. S1. Note that in the Supplementary Material, the convention for identifying the A and B atoms with boron and nitrogen is opposite to that used in Ref. [S7] and in the main text.

## VII. ADDITIONAL TIGHT-BINDING CALCULATIONS

We furthermore provide additional tight-binding calculations of optical responses within the effective model for twisted hexagonal boron nitride (t-hBN). For different stackings, we include (i) joint density of states (JDOS), (ii) first-order optical conductivities, (iii) optical weights, (iv) gauge-dependent shift vector and shift photoconductivity decompositions, (v) shift photoconductivities at different frequencies, (vi) spectral mean shift photoconductivities at different stackings, (vii) interlayer contributions and the spatial variations of the shift photoconductivities, (viii) integrated second-order shift photoconductivities, (ix) frequency-resolved and integrated second-order injection photoconductivities.

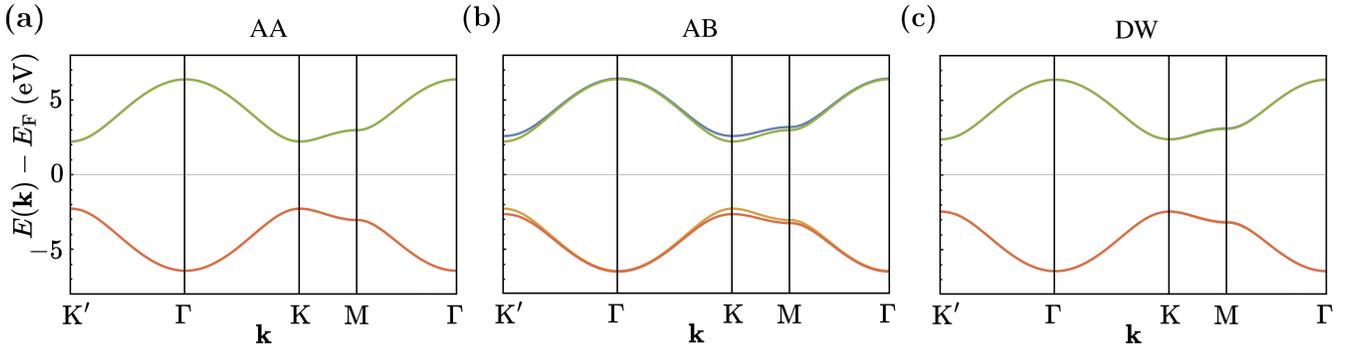

FIG. S1: Effective band structures obtained for the AA stacking **(a)**, AB stacking **(b)**, and DW stacking **(c)** configurations within the effective tight-binding model. The band structure quantitatively matches the *ab initio* band structure presented in the main text.

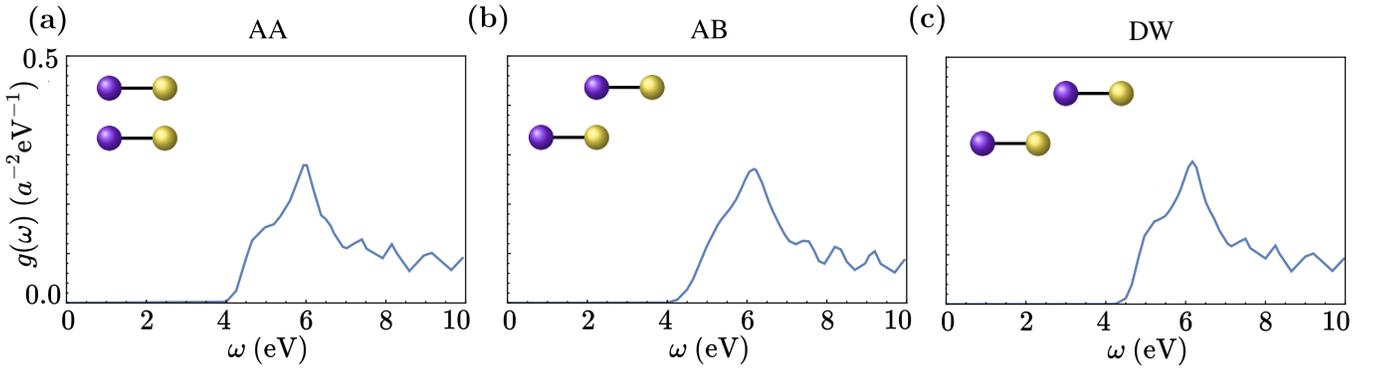

FIG. S2: The JDOS for optical transitions plotted as a function of the transition energy/$\Delta_0$ (where $\Delta_0 = 4.5\,\text{eV}$ is the band gap at the BZ edge). As expected, there is a flattening of the JDOS profile near the energy of the largest band gap ($\sim 5.3\,\text{eV}$) over the different stackings (responsible for the first peak in the photoconductivities) and a sharp peak near the saddle point energies ($\sim 6\,\text{eV}$) in the dispersion relation (responsible for the second peak in the photoconductivities).

### A. Joint density of states

In Fig. S2, we present numerical JDOS obtained within the model (see the previous Section) at the AA, AB, and DW (domain wall) stackings. We stress that in all cases, the dominant peak arises due to the resonant contributions from the M point of the Brillouin zone (BZ), cf. the main text, which correspond to the van Hove singularities. Another peak contribution arises at the frequency of the resonant transitions at the K point, as expected. While both characteristic points yield peak values in the JDOS, it should be emphasized that it is the response at the frequency resonant for the M point that is relevant for deducing the local polarization $\mathbf{P}(\mathbf{x})$. The latter is the consequence of the presence of dominant Berry connection $\mathbf{A}_{nn}$ contributions at the edges (rather than corners) of the BZ, which constitutes most of the electric polarizations evaluated within the configuration space approximation.

### B. First-order optical conductivity

Photocurrents in moiré hBN can be investigated by first employing Eq. (S8) to evaluate the first-order optical conductivities in the presence of an oscillating electric field.

A spectrally resolved calculation of first-order conductivity can be carried out using Eq. (S8) directly on replacing the $\delta$-function with a Gaussian of suitable width. To that end, the integration was carried out over a smoother $\mathbf{k}$-grid consisting of 469 $\mathbf{k}$-points to calculate the conductivity as a function of transition energy $\hbar\omega$. We show these in Figs. S3 (a-c) for the AA, AB and DW stackings. However, it must be noted that despite the smoother grids for numerical integration and the wide Gaussian approximations of the $\delta$-function, the curves are still not as smooth as would be expected from an analytical integration.

The qualitative nature of the curve is expected from a consideration of the competition between the transition rate $Q_{mn}^{ba}$ and the

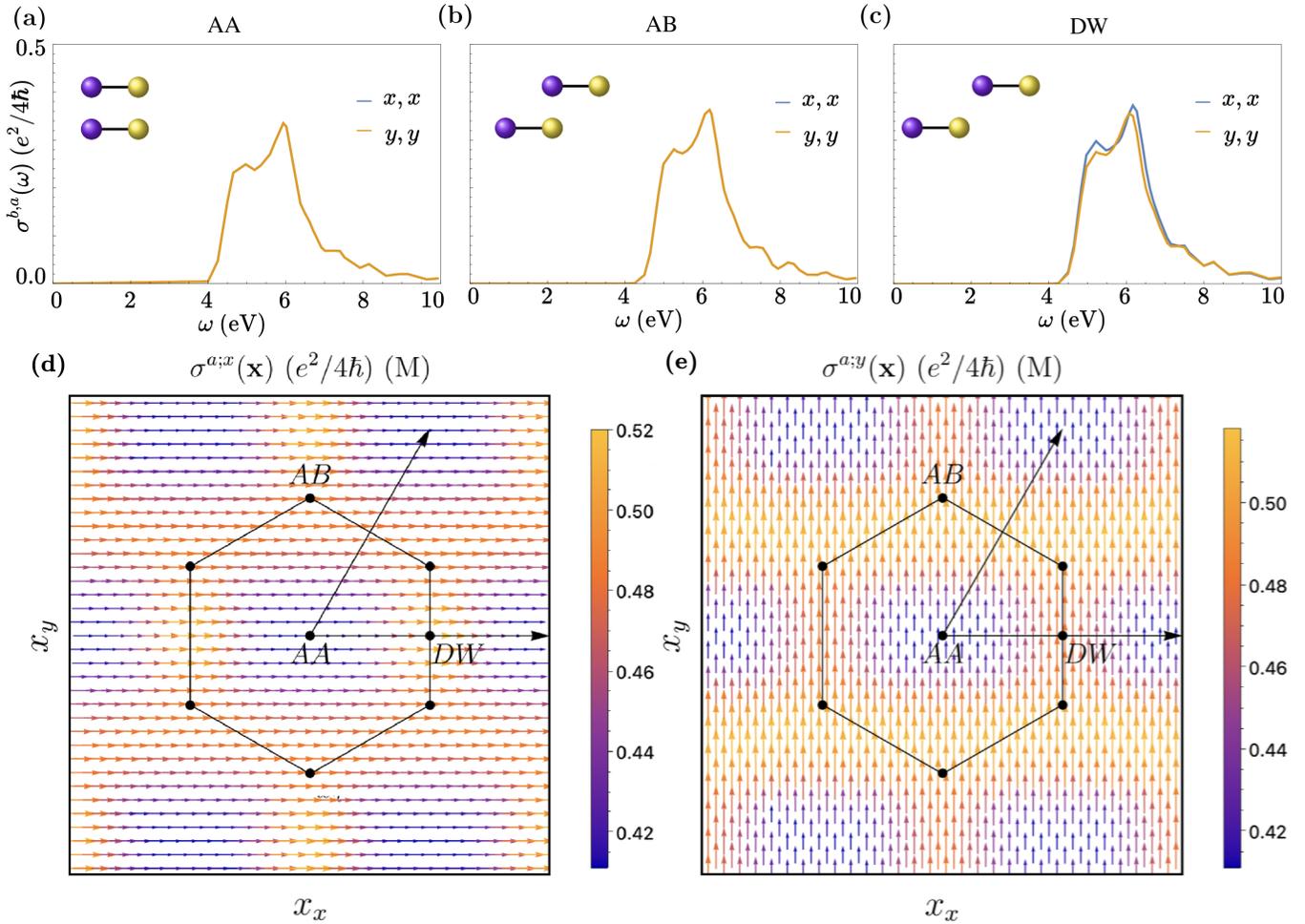

FIG. S3: **(a-c)** Spectrally resolved calculations of first-order photoconductivities $\sigma^{x,x}$ and $\sigma^{y,y}$ for stackings **(a)** AA, **(b)**, AB and **(c)** DW, as a function of optical transition energy/$\Delta_0$ (where $\Delta_0 = 4.5\,\text{eV}$ is the band gap at the BZ edge). The initial increase in conductivity can be attributed to an increase in the DOS that can undergo an optical transition at the relevant transition energy, while the subsequent decrease is due to both a decreasing optical transition rate ($\propto Q_{mn}^{ab}$) and a decreasing DOS. Such a qualitative trend matches the results from a superlattice potential model in [S16]. **(d-e)** First-order photocurrents associated with the optical conductivity as a function of stacking direction $\mathbf{x}$ in response to **(d)** $x$-polarized and **(e)** $y$-polarized light at frequency $\omega_{\text{M}} = 6\,\text{eV}$. The current direction is found to be approximately parallel to the polarization of the electric field, implying a negligible Hall response at first order. Also note that the qualitative dependence of the magnitude of the conductivity is similar to that of the optical weight in Figs. S4 (a,b).

density of states (DOS) $g(\omega_{mn})$. While the former decreases as the transition energy increases, being the highest at the edges of the BZ and decreasing upon moving towards the center [see Figs. S4(c), S4(d)], the DOS initially increases as transition energy increases, before finally decreasing as well. This suggests the qualitative form in Figs. S3 (a-c), where the conductivity initially increases from zero at the band gap owing to an increase in the DOS and then falls off after reaching a maximum – owing to a decrease both in the transition rate and the DOS. Such a spectrally resolved calculation of the conductivity has been performed for small transition energies in [S16] using a superlattice potential model to describe the moiré system and a Lorentzian to approximate the $\delta$-function; the general qualitative nature of the initially increasing conductivity matches their results. However, owing to the flatter bands associated with the superlattice potential model, the curve presented in [S16] is not as smooth as in Figs. S3 (a-c).

The vector plots of the first-order conductivity as a function of stacking in response to $x$- and $y$-polarized light at $\omega_{\text{M}} = 6\,\text{eV}$ are also shown in Figs. S3 (d-e). Importantly, the results show that the current direction is approximately the same as the electric field polarization, that is, there is no first order Hall response, and the qualitative dependence of the current magnitude on $\mathbf{x}$ is similar to that of the relevant optical weight [owing to the $Q_{mn}^{ab}$ factor in the integrand of Eq. (S8)]. Moreover, the current direction remains fairly uniform irrespective of the stacking. Since the stacking only enters the model via the (small)

interlayer hoppings, this suggests that the first-order conductivity is primarily an intralayer effect with the stacking-dependent interlayer effects only providing small variations of the order of $\sim 3.5\%$ on the background intralayer conductivity. This stands in contrast to the electronic polarization, which was observed to be a strong function of the stacking, an artifact of the property that monolayer hexagonal boron nitride is 3-fold symmetric, resulting in vanishing local polarization.

### C. Optical weights

To further study the optical conductivities in the context of this work, we define generalized **optical weights** following Ref. [S17] as frequency-weighted integrals of the first-order AC conductivities,

$$W_{ab}^{\alpha} \equiv \int_0^{\infty} d\omega \; \frac{\sigma^{b,a}(\omega)}{\omega^{\alpha}}. \tag{S29}$$

In the following, we choose $\alpha = 1$, which allows to target the QGT integrated over the entire BZ in the zero-temperature limit ($f_{nm} = 1$):

$$W_{ab}^1 = \int_0^{\infty} d\omega \; \frac{\sigma^{b,a}(\omega)}{\omega} = \frac{\pi e^2}{\hbar} \sum_{m,n} \int_{BZ} \frac{d^2\mathbf{k}}{(2\pi)^2} f_{nm} Q_{mn}^{ba}. \tag{S30}$$

Similarly to the approach for polarization above, the $\alpha = 1$ optical weight can be calculated as a function of $\mathbf{x}$ as a proxy of the optical absorption rate in the material. Using Eq. (S30) and numerically integrating over a $\mathbf{k}$-grid of 169 $\mathbf{k}$-points for 37 different stackings in the unit cell, $W_{xx}^1$ and $W_{yy}^1$ have been plotted in Figs. S4 (a,b). Note that owing to $W_{xy}^1$ being two orders of magnitude smaller than $W_{xx}^1$ and $W_{yy}^1$, it has been assumed negligible and not shown here.

Since the photoconductivities to be explored subsequently all involve factors of $Q_{mn}^{ab}$ in their integrands, it is useful to study its dependence on $\mathbf{k}$ in the BZ. Figures Figs. S4 (c,d) show this dependence for DW stacking; note that as with the trace of the QGT in the main text, $Q_{mn}^{ab}$ is largest at the edge of the BZ where the optical transition energy is the smallest. Importantly, the significance of the combination of the shift responses to $x$- and $y$-polarized light in $\sigma$ is further highlighted here: only by adding the two responses is the complete BZ edge and consequently, all of the dominant contribution to the polarization included in the shift current integral.

### D. Shift vectors and shift photoconductivity decomposition

In this Section, we demonstrate how the individual contributions of the shift vector decomposed in Eq. (S12) contribute to the shift photoconductivities. In particular, we show that by choosing an appropriate gauge, the last term can be made orders of magnitude smaller compared to the diagonal terms of non-Abelian Berry connection that contribute to the net electric polarization. We show the contributions of the individual terms over stacking in Fig. S5. In that context, we decompose the shift photoconductivities as $\sigma_{shift}^{c,ab}(\mathbf{x}) = \sigma_{\mathbf{P}}^{c,ab}(\mathbf{x}) + \sigma_{Arg}^{c,ab}(\mathbf{x})$, where band-diagonal Berry connection contribution to the shift photoconductivity reads: $\sigma_{\mathbf{P}}^{c,ab}(\mathbf{x})$ [Fig. S5(a)], whereas the contribution to the shift photoconductivities due to the non-Abelian Berry connection phase, i.e. the argument/phase/logarithm term reads: $\sigma_{Arg}^{c,ab}(\mathbf{x})$ [Fig. S5(b)]. As predicted, the argument term is an order of magnitude smaller than the polarization term and can be made vanishing in the optical gauge (not shown). In Fig. S5(c), we show the conduction band contributions to the first term resolved in $\mathbf{k}$-space. Notably, the conduction band contributions are equal and opposite to the valence band contributions, as featured in the main text. In Fig. S5(d), we show the difference of conduction and valence Berry connections entering the shift vectors ($R_{mn}^c \approx A_{mm}^c - A_{nn}^c$), which approximately equal twice the conduction band Berry connection contributions (corresponding to the negative of the electric polarization), as claimed in the considered moiré Hamiltonians.

It should be stressed again, as mentioned in the Methods, that under the optical gauge, the last (phase) term can be made vanishingly small. This allows to fully exploit and explore a direct connection between the shift photoconductivities and the electric polarization [S11, S18].

### E. Shift photoconductivities at different frequencies

In Fig. S6, we show that the shift photocurrent vortices associated with the local electric polarization can be observed in a broader frequency range $\omega$ (in twisted hBN, $\omega = 5.9 - 6.7$ eV). Therefore, the deduction of the electric polarization textures from the shift photocurrents is not limited to the resonant frequency $\omega_M = 6.0$ eV studied in the main text, and is robust for a

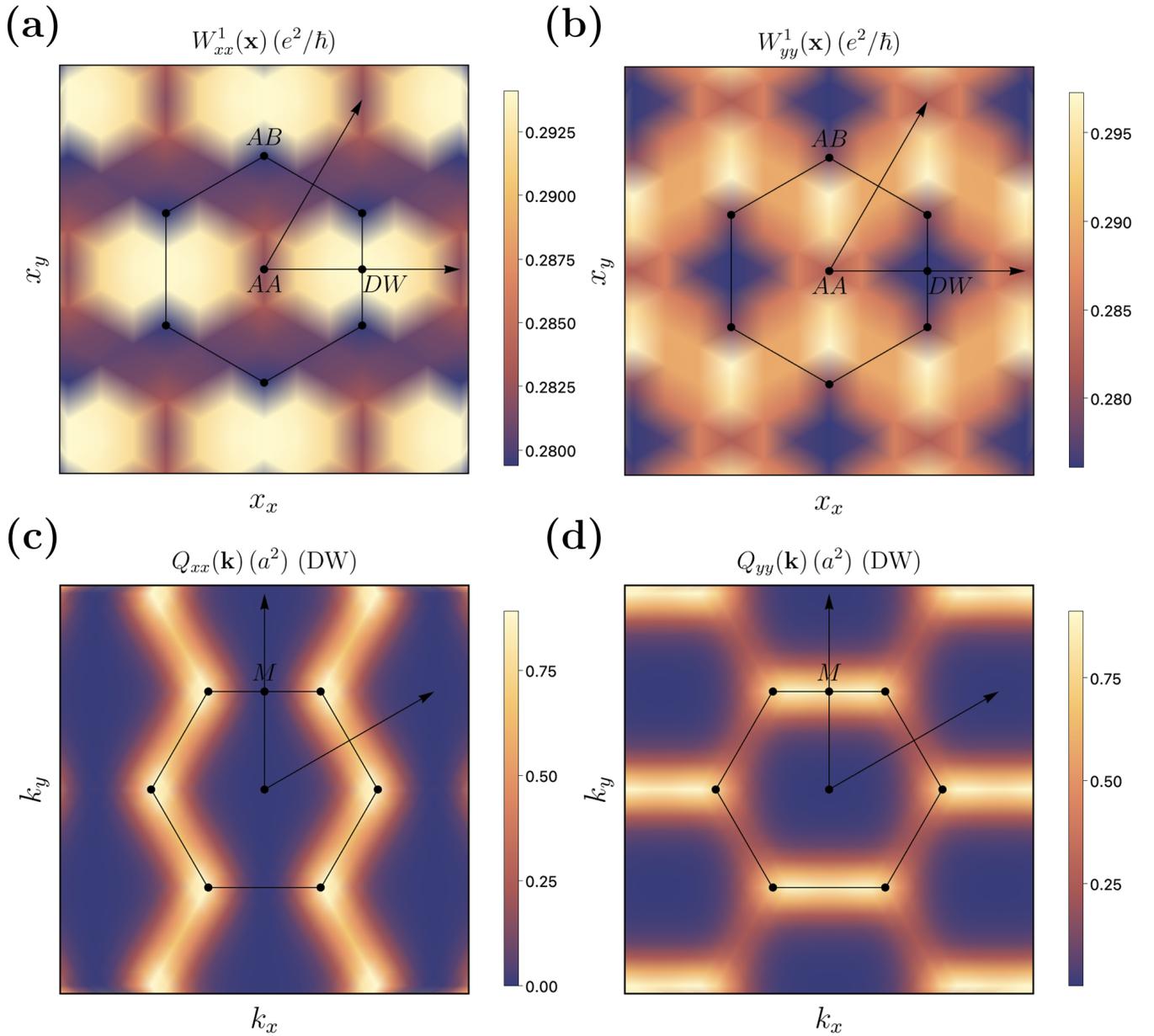

FIG. S4: **(a-b)** Optical weights for $\alpha = 1$ for an **(a)** $x$-polarized and **(b)** $y$-polarized driving electric field as a function of the stacking vector $\mathbf{x}$. $W^1_{xy}(\mathbf{x})$ is negligible compared to these contributions and is therefore not shown. **(c-d)** $Q^{xx}$ and $Q^{yy}$ summed over all valence-to-conduction interband transitions plotted over $\mathbf{k}$ in the BZ for DW stacking. As noted in the main text, adding the responses to $x$- and $y$-polarized light in $\boldsymbol{\sigma}$ ensures that $Q^{xx}$ and $Q^{yy}$ are added, and the entire BZ edge is probed.

lasing frequency window around the peak resonant frequency. Intuitively, the reason for the robustness is associated with the fact that these nearby frequencies still probe transitions in regions where the diagonal Berry connection is strong and contributes dominantly to both the electronic polarization and the shift photoconductivity.

### F. Spectral mean local shift photoconductivities

In Fig. S7, we demonstrate spectral mean local shift photoconductivities over different stacking. In the next subsection, we further contrast these findings with the spatially-resolved variations around the mean values. While the mean values are dominated by the intralayer contributions, we further attribute the variations to the interlayer contributions that are induced by the relative bilayer stacking.

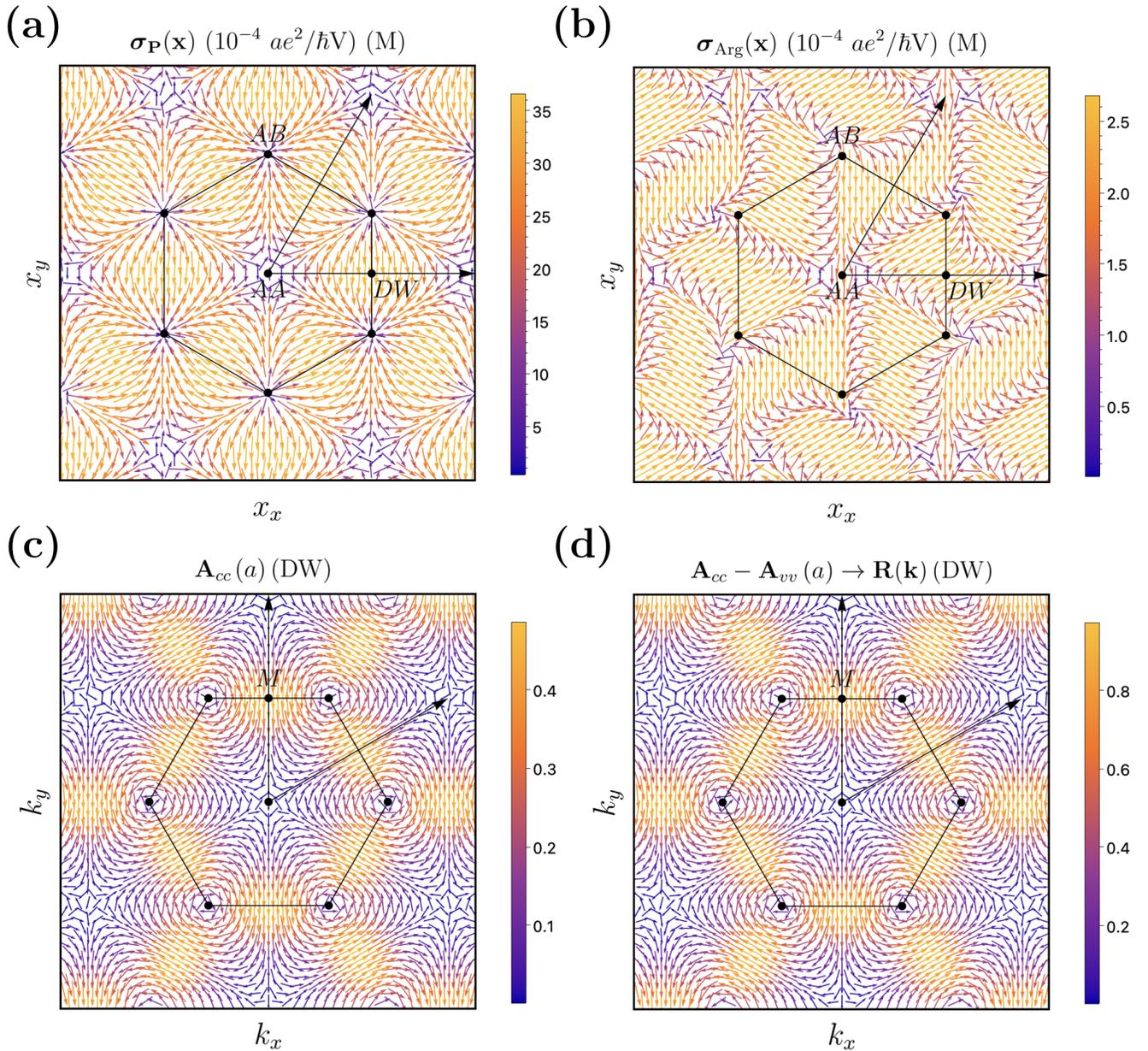

FIG. S5: **(a)** Band-diagonal Berry connection contribution to the shift photoconductivity vector, $\boldsymbol{\sigma}_{\mathbf{P}}(\mathbf{x})$, which reflects the change of electric polarization on interband optical transitions. **(b)** Contribution to the shift photoconductivities due to the non-Abelian Berry connection phase, i.e. the argument (Arg) term, $\boldsymbol{\sigma}_{\mathrm{Arg}}(\mathbf{x})$. We observe that under the chosen fixed gauge, the latter is an order of magnitude smaller [$\sigma_{\mathbf{P}}(\mathbf{x}) \gg \sigma_{\mathrm{Arg}}(\mathbf{x})$]. **(c)** Berry connection in the conduction bands, $\mathbf{A}_{cc}(\mathbf{k})$. In the main text, the Berry connection in the valence band, contributing to both the local polarization and local shift photoconductivities, was presented. We observe that in the moiré Hamiltonian defining the effective model, manifestly: $\mathbf{A}_{cc}(\mathbf{k}) \approx -\mathbf{A}_{vv}(\mathbf{k})$. **(d)** Differences of Berry connections in conduction and valence bands entering the shift vector $\mathbf{R}(\mathbf{k})$. We observe that the shift vector magnitude $|\mathbf{R}(\mathbf{k})|$ doubles the value of $|\mathbf{A}_{vv}(\mathbf{k})|$ across the BZ, consistently with the model introduced in Methods.

As earlier with the first-order response, the homogeneity of the shift current direction with stacking vector indicates that intralayer effects dominate, an outcome that is expected since an optical transition from the valence to conduction bands effectively involves the Wannier center shifting from the near the N atoms to near the B atoms, a non-zero change even within a single layer.

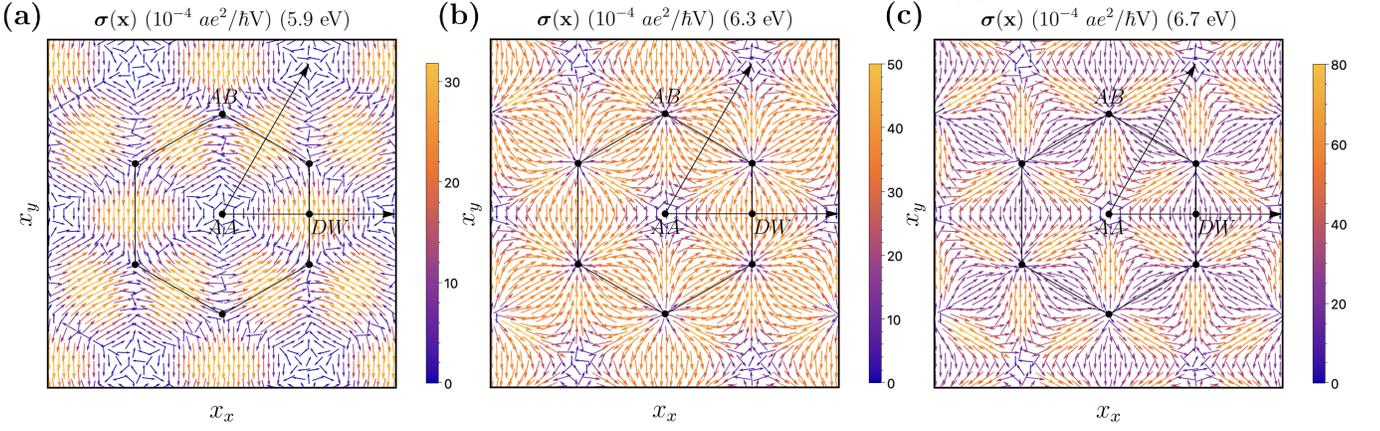

FIG. S6: Shift photoconductivity vector $\boldsymbol{\sigma}(\mathbf{x})$ in twisted boron nitride (t-hBN) plotted over stacking at different frequencies: **(a)** $\omega = 5.9\,\text{eV}$, **(b)** $\omega = 6.3\,\text{eV}$, and **(c)** $\omega = 6.7\,\text{eV}$. As in the main text, where we analogously demonstrate $\boldsymbol{\sigma}(\mathbf{r})$ at the resonant frequency $\omega_M = 6.0\,\text{eV}$, the shift photoconductivity vectors $\boldsymbol{\sigma}$ here too are antiparallel to the in-plane polarization. We find that this feature is robust within an extended frequency range $\omega = 5.9 - 6.7\,\text{eV}$, showing that a lasing frequency window, rather than a single resonant frequency, is admitted by the effect. As might be expected, at the edges of this frequency window [as in **(c)**], while the direction remains anti-parallel to the polarization, the magnitudes are no longer reliable.

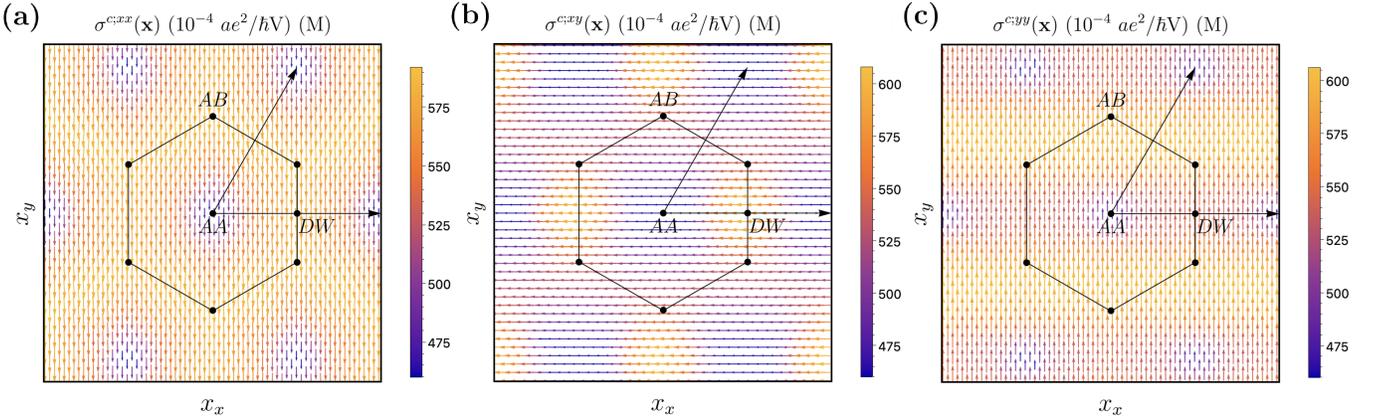

FIG. S7: Spectral mean shift photoconductivities over different stackings in response to **(a)** $x$-polarized, **(b)** $xy$-mixed, and **(c)** $y$-polarized electric field stimuli. The variations around the mean shift photoconductivities values are shown in Fig. S8. The shift photoconductivities respect the crystalline symmetries of the bilayer subject to the oscillating electric fields.

## G. Interlayer contributions and the variation of local shift photoconductivities

Having established the uniformity in direction in the various shift photoconductivity components, it is interesting to investigate the deviation of the shift current response from the mean. More precisely, $\delta\sigma^{c,ab}_{\text{shift}} = \sigma^{c,ab}_{\text{shift}} - \int \frac{\mathrm{d}^2 x}{A_{\text{cell}}}\sigma^{c,ab}_{\text{shift}}$, where $A_{\text{cell}}$ is the area of the unit cell in real space. This has been plotted for the $xx$, $xy$, and $yy$ electric field stimuli in Fig. S8. As expected, these deviations show a strong dependence, both in magnitude and direction, on $\mathbf{x}$. Note that the plots in response to the $xx$ and $yy$ stimuli show a winding across the unit cell similar to the polarization textures shown in the main text.

The variations of the the local shift photoconductivities can be attributed to the interlayer contributions induced on stacking two twisted layers, see Fig. S8. It should be stressed that these terms vanish completely in the limit of (infinitely) separated monolayers.

## H. Integrated shift photoconductivities

Furthermore, the second-order optical response Eq. (S11) can be used to calculate the total integrated shift photoconductivity. To that end, the responses to 'white light' were found (that is, Eq. (S11) was integrated over $\omega$ to remove the $\delta$-function), and the

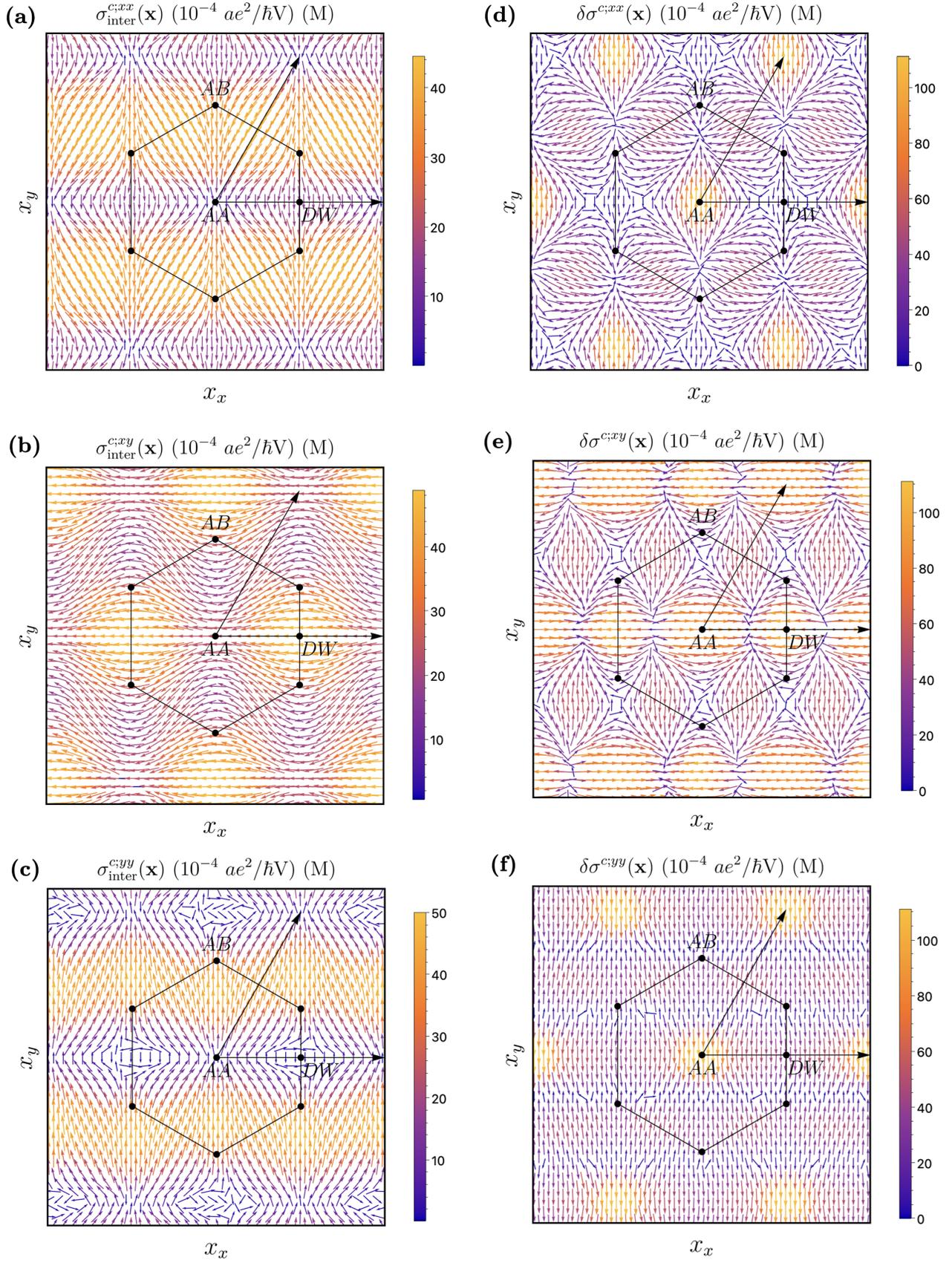

FIG. S8: Interlayer contributions to local shift photoconductivities **(a-c)**, and deviations from the mean of the shift photoconductivities **(d-f)** as a function of stacking direction in response to: **(a,d)** $xx$, **(b,e)** $xy$, and **(c,f)** $yy$ electric field stimuli.

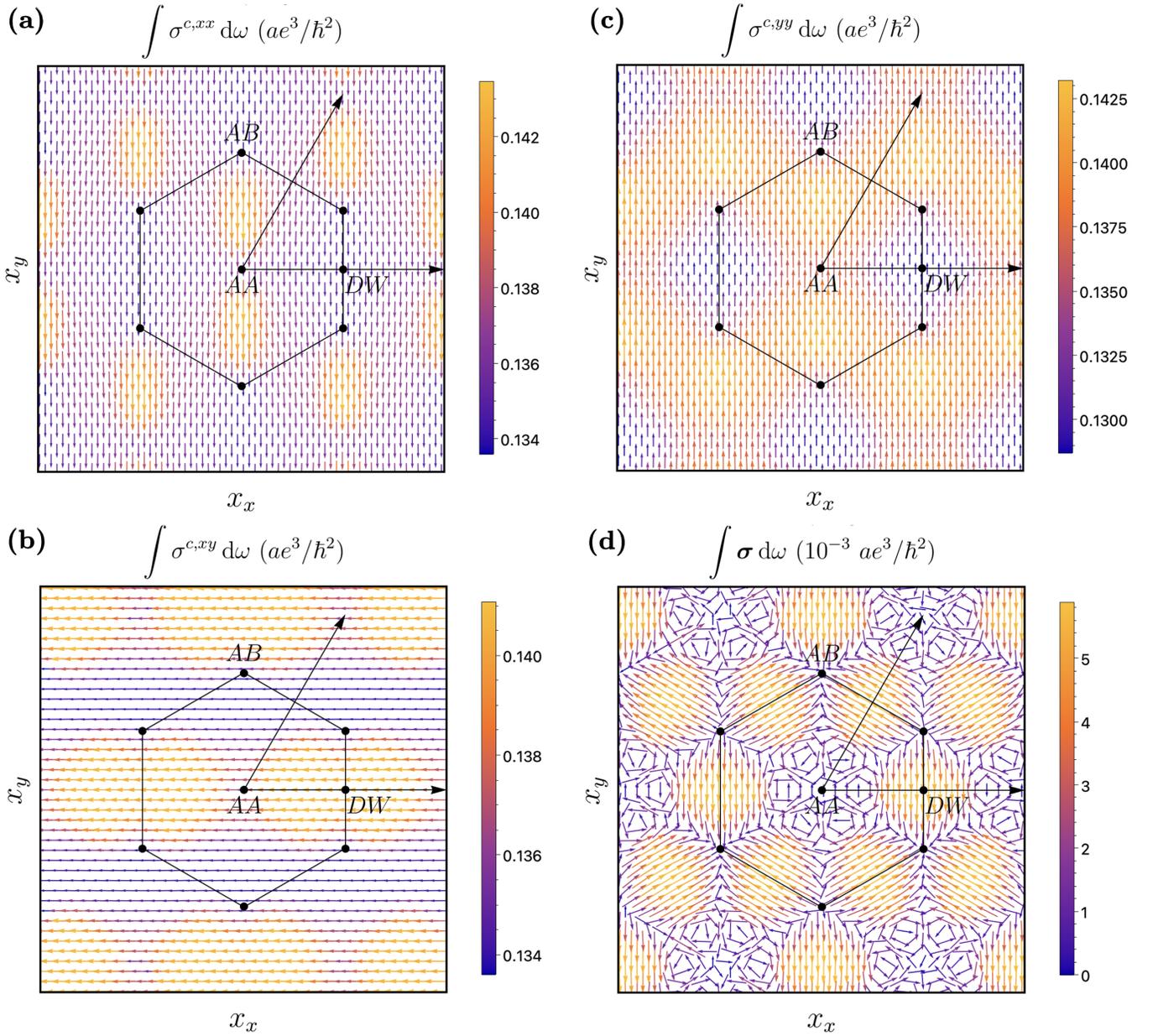

FIG. S9: **(a-c)** Integrated shift photoconductivities as a function of stacking, in response to **(a)** $xx$, **(b)** $xy$, and **(c)** $yy$ electric field stimuli. As with the first-order conductivities in Figs. S3 (d-e), an approximately uniform current direction over stacking implies dominant intralayer effects. **(d)** Summed shift photocurrents $\boldsymbol{\sigma}(\mathbf{r})$ in a vectorized form, cf. main text, as a function of stacking. This is antiparallel to the in-plane polarization texture as well (except near AA) and can also serve as a tool for probing the polarization in addition to measuring $\boldsymbol{\sigma}$ near $\omega_M$.

numerical integration in the BZ was carried out over a $\mathbf{k}$-grid consisting of 169 $\mathbf{k}$-points for 37 different stackings. The results are presented in Fig. S9 as separate vector plots over $\mathbf{x}$ for the current direction in response to $xx$, $xy$, and $yy$ stimuli.

Importantly, Fig. S9 (d) presents the integrated shift photoconductivity vector $\boldsymbol{\sigma}$ as a function of stacking and shows that in addition to shift responses measured near $\omega_M$, an integrated response is also antiparallel to the electronic polarization texture. Hence, it can equally be used as a probe for the same deductions. However, care must be taken to ensure that such an integrated shift method is not relied upon at small polarization magnitudes, as evidenced by the behaviour of $\int \boldsymbol{\sigma} \, \mathrm{d}\omega$ near the AA stacking.

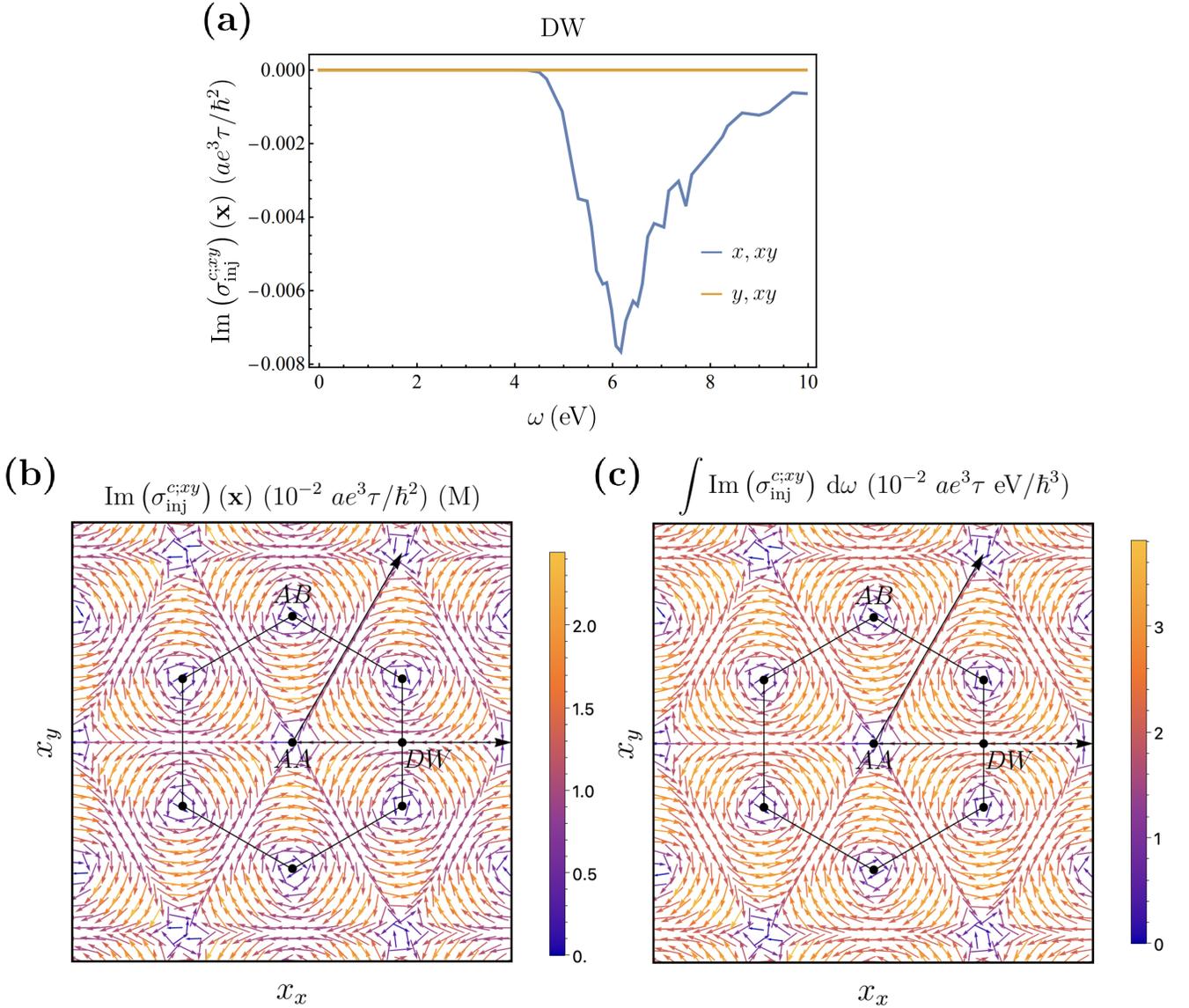

FIG. S10: **(a)** Frequency-resolved injection photoconductivities at the DW stacking in response to $xy$ electric field stimuli. **(b)** Injection photoconductivities at the M point resonant frequency $\omega_M$, as a function of stacking in response to an $xy$ electric field. The $\sigma_{\mathrm{inj}}^{y,xy}$ component vanishes at the DW consistently with the spectrally resolved photoconductivities in panel **(a)**. **(c)** Integrated injection photoconductivities as a function of stacking in response to an $xy$ electric field stimulus. $\sigma_{\mathrm{inj}}^{c,xx}$ and $\sigma_{\mathrm{inj}}^{c,yy}$ are negligible and hence not shown here. Importantly, such a vanishing of $\sigma_{\mathrm{inj}}^{c,xx}$ and $\sigma_{\mathrm{inj}}^{c,yy}$ increases the experimental accessibility to the electric polarization-related $\sigma_{\mathrm{shift}}^{c,xx}$ and $\sigma_{\mathrm{shift}}^{c,yy}$. Additionally, upon a comparison of the units in these plots (see the text), it can be concluded that for mean-free-time $\tau \sim 10^{-15}$ s, injection currents are an order of magnitude smaller than the shift ones, deeming these unlikely to be visible in an experiment.

## I. Second-order injection photocurrents

At second order in the driving electric field, Eq. (S10) can be used to calculate the injection photoconductivities. Notably, the magnitudes of the injection responses realized in t-hBN to linearly polarized light are negligibly small and should not interfere with measurements of $\boldsymbol{\sigma}$, as mentioned in the main text. For both frequency-resolved injection photoconductivities and vector plots of injection responses as a function of stacking, see Fig. S10. Contrary to first-order and shift responses, the strong dependence of the injection currents on $\mathbf{x}$ suggests that the intralayer contributions to injection currents are negligible and the interlayer effects dominate.

The ratio of the units between the shift and injection current magnitudes is $1 : \frac{2eV}{\hbar}\tau \sim 1 : 1$ (for $\tau \sim 10^{-15}$ s), which implies that injection effects are an order of magnitude smaller than shifts and are likely to not be significant in experimental observations of second-order effects. Furthermore, owing to the dependence of the injection currents on mean-free-time $\tau$, it is possible to tune them out by employing experimental methods that decrease $\tau$ such as increasing the temperature and thus increasing the number of activated phonon modes in the phonon-mediated resistivity regime, or by increasing the defect concentration in the defect-mediated resistivity regime.

[S1]  J. Provost and G. Vallee, Commun. Math. Phys. **76**, 289 (1980).
[S2]  J. Ahn, G.-Y. Guo,  and N. Nagaosa, Phys. Rev. X **10**, 041041 (2020).
[S3]  J. Ahn, G.-Y. Guo, N. Nagaosa,  and A. Vishwanath, Nat. Phys. **18**, 290 (2021).
[S4]  J. E. Sipe and E. Ghahramani, Phys. Rev. B **48**, 11705 (1993).
[S5]  J. E. Sipe and A. I. Shkrebtii, Phys. Rev. B **61**, 5337 (2000).
[S6]  R. King-Smith and D. Vanderbilt, Phys. Rev. B **47**, 1651 (1993).
[S7]  D. Bennett, G. Chaudhary, R.-J. Slager, E. Bousquet,  and P. Ghosez, Nat. Commun. **14**, 1629 (2023).
[S8]  D. Bennett, W. J. Jankowski, G. Chaudhary, E. Kaxiras,  and R.-J. Slager, Phys. Rev. Res. **5**, 033216 (2023).
[S9]  W. J. Jankowski, D. Bennett, A. Agarwal, G. Chaudhary,  and R.-J. Slager, arXiv:2404.16919  (2024).
[S10] S. Carr, D. Massatt, S. B. Torrisi, P. Cazeaux, M. Luskin,  and E. Kaxiras, Phys. Rev. B **98**, 224102 (2018).
[S11] B. M. Fregoso, T. Morimoto,  and J. E. Moore, Phys. Rev. B **96**, 075421 (2017).
[S12] D. Vanderbilt and R. King-Smith, Phys. Rev. B **48**, 4442 (1993).
[S13] C. Hu, M. H. Naik, Y.-H. Chan, J. Ruan,  and S. G. Louie, PNAS **120**, e2314775120 (2023).
[S14] H. Yu, Z. Zhou,  and W. Yao, Sci. China Phys. Mech. Astron. **66**, 107711 (2023).
[S15] A. H. Castro Neto, F. Guinea, N. M. R. Peres, K. S. Novoselov,  and A. K. Geim, Rev. Mod. Phys. **81**, 109 (2009).
[S16] H. Ochoa and A. Asenjo-Garcia, Phys. Rev. Lett. **125**, 037402 (2020).
[S17] Y. Onishi and L. Fu, Physical Review X **14** (2024), 10.1103/physrevx.14.011052.
[S18] R. Resta, arXiv:2402.12489  (2024).


\clearpage

\end{document}